\documentclass[aps,prb,amsmath,amsfonts,amssymb]{revtex4}

\usepackage{graphicx}
\usepackage{epsfig}

\def\de{{\rm d}}
\def\s{\sigma}
\def\us{\underline{\sigma}}
\def\Ns{{N_{\rm s}}}
\def\Nt{{{\cal N}_{\rm traj}}}
\def\bs{{\boldsymbol{\sigma}}}
\def\ubs{\underline{\boldsymbol{\sigma}}}
\def\hE{\widehat{E}}
\def\hH{\widehat{H}}
\def\hO{\widehat{O}}

\def\tE{\widetilde{E}}
\def\Tr{{\rm Tr}}
\def\ve{\varepsilon}
\def\tve{\widetilde{\varepsilon}}
\def\tq{\widetilde{q}}
\def\bu{{\boldsymbol{u}}}
\def\bh{{\boldsymbol{h}}}
\def\ba{{\boldsymbol{a}}}
\def\bb{{\boldsymbol{b}}}
\def\tv{\widetilde{v}}
\def\eqd{\overset{\rm d}{=}}
\def\E{\mathbb{E}}
\def\I{{\rm 1\hspace{-0.90ex}1}}

\def\Z{{\cal Z}}

\begin{document}
\date{\today}

\title{On the path integral representation for quantum spin models and its
  application to the quantum cavity method and to Monte Carlo simulations}

\author{Florent Krzakala}
\affiliation{Laboratoire PCT, Unit\'e Mixte
  de Recherche (UMR 7083 Gulliver) du CNRS et de l'ESPCI
  ParisTech, 10 rue Vauquelin, 75231 Paris, France}
\author{Alberto Rosso}
\affiliation{LPTMS, 
  Unit\'e Mixte de Recherche (UMR8626)
  du CNRS et de l'Universit\'e Paris-Sud,
  B\^at.~100, Universit\'e Paris-Sud, 91405 Orsay
  Cedex, France}
\author{Guilhem Semerjian}
\author{Francesco Zamponi}
\affiliation{LPTENS, Unit\'e Mixte de Recherche (UMR 8549) du CNRS et
  de l'ENS, associ\'ee \`a l'UPMC Univ Paris 06, 24 Rue Lhomond, 75231
  Paris Cedex 05, France.}

\begin{abstract}
The cavity method is a well established technique for solving classical spin
models on sparse random graphs (mean-field models with finite connectivity). 
Laumann, Scardicchio and Sondhi [{\tt arXiv:0706.4391}] proposed recently an extension of this method to quantum 
spin-1/2 models in a transverse field, using a discretized Suzuki-Trotter imaginary
time formalism. Here we show how
to take analytically the continuous imaginary time limit. Our main technical
contribution is an explicit procedure to generate the spin trajectories in a
path integral representation of the imaginary time dynamics. As a side result
we also show how this procedure can be used in simple heat-bath like
Monte Carlo simulations of generic quantum spin models. The replica symmetric
continuous time quantum cavity method is formulated for a wide class of
models, and applied as a simple example on the Bethe lattice ferromagnet in a
transverse field. The results of the methods are confronted with various
approximation schemes in this particular case. On this system we performed
quantum Monte Carlo simulations that confirm the exactness of the cavity
method in the thermodynamic limit.
\end{abstract}

\maketitle

\section{Introduction}

Mean-field approximations are often useful first steps to unveil the physical
content of realistic models. This is all the more true when
exact solutions are probably impossible to obtain in the finite-dimensional 
setting, in particular when quenched disorder and/or quantum effects have to be
taken into account, as for instance in the case of Anderson 
localization~\cite{popu_first}.
Another example is dynamical mean-field 
theory~\cite{DMFT}, that has been a very fertile approach to the problem of strongly
correlated fermions. It can be sometimes preferable to study mean-field
theories not by making an approximation to a finite-dimensional model, but
rather by formulating a model which is mean-field by nature, this allowing in
particular to state the results in a mathematically clearer way. The simplest
of such examples is the Curie-Weiss model of ferromagnetism, in which $N$
classical Ising spins all interact attractively 
which each other, with a coupling constant
scaling inversely with the size of the system to ensure a well-defined
thermodynamic limit. The equivalent for quenched disordered systems is the
Sherrington-Kirkpatrick model~\cite{SK} of a spin glass, where again all spins
interact weakly with other, yet with coupling constants of random signs.

The mean-field character of the above mentioned models arises from their
infinite connectivity (in the thermodynamic limit). There exists however
another class of models, which are still mean-field yet keep a
finite connectivity, each of the degrees of freedom they possess interacting
only with a finite (with respect to $N$) number of neighbors.
For ferromagnetic models they can be obtained by the Cayley tree construction,
where one draws an infinite regular tree and studies the magnetization of the
root site~\cite{Baxter}. Cayley tree models have however pathological surface
effects, and the theory of finitely-connected mean-field frustrated systems is
better defined on random graphs~\cite{VB,MePa}, of fixed or fluctuating
connectivity. 
Classical models of spins on such random structures have been the subject of
extensive study in the last decade. These works were motivated on the one hand
by their somehow more physically realistic features, namely the finite
connectivity, and also because of their strong relationship with issues
originating from computer science, namely the understanding of phase
transitions in random constraint satisfaction 
problems~\cite{MoZe,MePaZe,pnas}. 
Finite-connectivity
models are technically much more involved than their fully-connected
counterparts. The replica method~\cite{Beyond} that has been first developed
to solve the 
Sherrington-Kirkpatrick model becomes less practical in this
setting~\cite{replica_diluted}, and the 
alternative cavity method turned out to be more useful~\cite{MePa}.

The interplay between quenched disorder and quantum fluctuations can lead to
a very rich phenomenology, and in particular the properties of the glass phase
found at low temperatures in classical models can be qualitatively modified
when a transverse field acts on the system~\cite{Giulio}. 
More generally the issue of the 
nature of the quantum phase transitions at zero temperature~\cite{Sachdev} in
presence of disorder is a very rich one. In the context of mean-field theory
this point has been mainly studied in fully-connected
models~\cite{BrayMooreQSK,Goldschmidt,Nish_stat,Georges,Leticia,Chamon01,MI08}, with a few
exceptions that appeared in the last year~\cite{BT07,qc_first,qksat,qbl}. 
In a very interesting
contribution~\cite{qc_first} Laumann, Scardicchio and Sondhi made a first 
step in extending the
cavity method to quantum spin models in a transverse field, and in this paper
we shall develop further this idea by solving a discretization problem which
plagued their proposal. Let us also mention here the work of Knysh and 
Smelyanskiy~\cite{qksat}, who developed a similar approach in the framework 
of the so-called static 
approximation~\cite{BrayMooreQSK,Nish_stat,Leticia}.
The motivations for this line of work is two-fold.
On the one hand one can expect an even richer physical behavior of
finitely-connected quantum models with respect to the fully-connected
ones. The possible fluctuations in the local geometry, and some notion of
distance which was absent in fully-connected models opens the way to a more
complex phenomenology. On the other hand one can aim at a better understanding
of some issues of quantum computing, and in particular on the use of quantum
annealing (or adiabatic algorithm)~\cite{diego,annealing,adiabatic} 
to solve random
constraint satisfaction problems. These quantum algorithms do indeed rely on
the application of a transverse field on spin models that have been
extensively studied at the classical level with the cavity method. Computing
the location and nature of the phase transitions~\cite{qrem} 
encountered along the
annealing path (as the transverse field is progressively turned off) might
give some informations on the behaviour of these quantum algorithms themselves.

The remaining of the article is organized as follows.
In Sec.~\ref{sec_st_ct} we recall the Suzuki-Trotter approach to spin-1/2
models in a transverse field, and develop our main technical contribution in
Sec.~\ref{sec_ct}, where we show how to actually build the spin trajectories
of the path integral representation of the imaginary time evolution operator.
Section~\ref{sec_qbl} is then devoted to the study of a very simple example of
finitely-connected quantum model, namely the Bethe lattice ferromagnet. We
first explain the continuous time quantum cavity treatment of this model,
before presenting the results of the method and confronting them with some
approximate approaches. In Sec.~\ref{sec_qmc} we present numerical results of
Monte Carlo simulations we performed for this model, and show how the
computations of Sec.~\ref{sec_ct} can be turned in a simple and versatile
quantum Monte Carlo method. The generic formalism of the quantum cavity method
is developed in Section~\ref{sec_qcav_generic}; we hope this order of
presentation, and the inclusion of a fully worked-out example before the
general case, will ease the reading of this work. We finally draw our
conclusions and put forward perspectives for future work in
Sec.~\ref{sec_conclusions}. Some technical details are deferred to a series of
Appendices.

\section{Path integral representation for spin-1/2 models}
\label{sec_st_ct}

\subsection{Spin models in a transverse field: Suzuki-Trotter formalism}
\label{sec_st}
Let us consider the Hilbert space spanned by the orthonormal basis of $2^N$
kets $|\us\rangle$, where $\us=(\s_1,\dots,\s_N)$ denotes a configuration of
$N$ Ising spins, $\s_i = \pm 1$. This space can be viewed as the tensorial
product of $N$ spins $1/2$, with operators $\s_i^z$ and $\s_i^x$, whose action
on the base vectors is defined by
\begin{eqnarray}
\s_i^z |\us\rangle &=& \s_i |\us\rangle \ , \ \\ 
\s_i^x |\us\rangle &=& |\s_1,\dots,\s_{i-1},-\s_i,\s_{i+1},\dots,\s_N\rangle \ .
\nonumber
\end{eqnarray}
From a classical energy function of $N$ Ising spins, $E(\s_1,\dots,\s_N)$, one
can construct an operator $\hE=E(\s_1^z,\dots,\s_N^z)$, diagonal in the 
$\{|\us\rangle\}$ basis. 
The Hamiltonian operators investigated in this paper are
obtained from such a classical energy by the addition of a transverse field,
\begin{equation}
\hH = \hE - B \sum_{i=1}^N \s_i^x \ , \quad \text{with} \ B \ge 0 \ .
\end{equation}
Our goal is then to compute the quantum statistical mechanics properties at
inverse temperature $\beta$, i.e. the partition function $Z$ and the average 
of observables (operators) $\hO$, defined by
\begin{equation}
Z=\Tr \left( e^{-\beta \hH}  \right) \ , \qquad 
\langle \hO \rangle = \frac{\Tr \left( \hO \ e^{-\beta \hH}  \right) }
{\Tr \left( e^{-\beta \hH}  \right) } \ .
\end{equation}
A well-known way of tackling such problems is to transform them into an
extended Ising model by using the Suzuki-Trotter formula \cite{suzuki}, as summarized in the
following lines :
\begin{eqnarray}
Z &=& \Tr \left(\left( e^{-\frac{\beta}{\Ns} \hE + 
\frac{\beta}{\Ns}B \sum_{i=1}^N \s_i^x  } \right)^\Ns \right) \\
&=& \lim_{\Ns \to \infty } \Tr \left(\left( e^{-\frac{\beta}{\Ns} \hE} 
e^{\frac{\beta}{\Ns}B \sum_{i=1}^N \s_i^x  } \right)^\Ns \right) \nonumber \\
&=& \lim_{\Ns \to \infty } \sum_{\us^1 , \dots , \us^\Ns}
\prod_{\alpha=1}^\Ns \ \langle \us^\alpha | e^{-\frac{\beta}{\Ns} \hE}
\ e^{\frac{\beta}{\Ns}B \sum_{i=1}^N \s_i^x  }  | \us^{\alpha+1} \rangle \nonumber\\
&=& \lim_{\Ns \to \infty } \sum_{\us^1 , \dots , \us^\Ns}
\prod_{\alpha=1}^\Ns e^{-\frac{\beta}{\Ns} E(\us^\alpha)} 
\prod_{i,\alpha}
\langle \s_i^\alpha | e^{\frac{\beta}{\Ns} B \s^x} | \s_i^{\alpha+1} \rangle 
\ .\nonumber
\end{eqnarray}
In the two last lines $\us^{\Ns+1} = \us^1$.
For a finite value of the number of Suzuki-Trotter ``slices'' $\Ns$, the
problem has thus become one of $N\times \Ns$ Ising spins, each of the 
$\s_i$ being promoted to a ring $(\s_i^1,\dots,\s_i^\Ns)$ with nearest
neighbor ferromagnetic interactions along the ``discrete imaginary time''
$\alpha$ axis (with periodic boundary conditions). The original 
interactions $E$ acts indentically and independently on each of the
configurations $\us^\alpha$. For notational convenience we shall use
bold symbols for quantities that depend on the slice $\alpha$, for instance
$\bs_i=(\s_i^1,\dots,\s_i^\Ns)$ is the configuration of the ring of Ising 
spins at site $i$ and $\ubs=(\us^1,\dots,\us^\Ns)$ is the full configuration
of the $N \times \Ns$ spins. We can
thus introduce a probability measure on the $N\times \Ns$ Ising spins,
\begin{eqnarray}
\mu(\ubs) &=& \frac{1}{Z_\Ns} e^{-\beta \tE(\ubs)} \prod_{i=1}^N w(\bs_i) \ , 
\label{eq_def_mu_ubs} \\
Z_\Ns &=& \sum_{\ubs} e^{-\beta \tE(\ubs)} 
\prod_{i=1}^N w(\bs_i) \ ,
\nonumber
\end{eqnarray}
such that the normalization constant $Z_\Ns$ reduces to the partition 
function $Z$ in the $\Ns\to\infty$ limit (in the following we shall sometimes
keep implicit the dependence on $\Ns$). To write in a compact way this last 
equation we have defined
\begin{equation}
\tE(\ubs) = \frac{1}{\Ns} \sum_{\alpha=1}^\Ns E(\us^\alpha) \ ,
\end{equation}
the average of independent copies of the classical energy on the various slices, and
\begin{widetext}
\begin{equation}
w(\bs) = \prod_{\alpha=1}^\Ns 
\langle \s^\alpha | e^{\frac{\beta}{\Ns} B \s^x} | \s^{\alpha+1} \rangle 
= \prod_{\alpha=1}^\Ns \left( \cosh\left(\frac{\beta B}{\Ns} \right)
  \delta_{\s^\alpha , \s^{\alpha +1}} + \sinh\left(\frac{\beta B}{\Ns}\right)
  \delta_{\s^\alpha , - \s^{\alpha +1}}   \right)
\ , \label{eq_def_w}
\end{equation}
\end{widetext}
the ferromagnetic interaction along the imaginary time axis
induced by the transverse field (we use 
$\s^{\Ns+1} = \s^1$). One can easily show that the average value of
observables can be obtained in this formalism as
\begin{equation}
\langle \hO \rangle = \sum_{\ubs} \mu(\ubs)
\frac{\langle \us^\alpha | \hO e^{-\frac{\beta}{\Ns}\hH}|\us^{\alpha+1}\rangle}
{\langle \us^\alpha | e^{-\frac{\beta}{\Ns}\hH}|\us^{\alpha+1}\rangle} \ ,
\end{equation}
where the slice number $\alpha$ is here arbitrary, thanks to the cyclic 
invariance around the discrete imaginary time axis. This can be simplified
further for observables $\hO$ diagonal in the $\{|\us\rangle \}$ basis, 
i.e. that can be written as $O(\s_1^z,\dots,\s_N^z)$:
\begin{equation}
\langle \hO \rangle = \sum_{\ubs} \mu(\ubs) O(\us^\alpha)
=  \sum_{\ubs} \mu(\ubs) 
\frac{1}{\Ns} \sum_{\alpha=1}^\Ns O(\us^\alpha) \ .
\label{eq_obs_diag}
\end{equation}
A non-diagonal observable we shall study in the following is the transverse
magnetization $\langle \s_i^x \rangle$ (written here for an arbitrary site 
$i$), which can be computed as
\begin{eqnarray}
\langle \s_i^x \rangle &=&  \sum_{\ubs} \mu(\ubs) 
\frac{1}{\Ns} \sum_{\alpha=1}^\Ns 
\frac{\langle \s_i^\alpha | \s^x e^{\frac{\beta}{\Ns} B \s^x} | 
\s_i^{\alpha+1}\rangle}
{\langle \s_i^\alpha | e^{\frac{\beta}{\Ns} B \s^x} | \s_i^{\alpha+1}\rangle} 
\label{eq_mx_nbfinite} =  \sum_{\ubs} \mu(\ubs) 
\frac{1}{\Ns} \sum_{\alpha=1}^\Ns 
\left(\tanh\left(\frac{\beta B}{\Ns} \right) 
\right)^{\s_i^\alpha \s_i^{\alpha+1}} \ .
\nonumber
\end{eqnarray}

\subsection{The continuous imaginary time limit}
\label{sec_ct}

To recover the truly quantum properties of the model one has to perform the
limit $\Ns \to \infty$. The basic degrees of freedom $\bs_i$ which were the 
configurations of a ring of Ising spins $(\s_i^1,\dots,\s_i^\Ns)$ then 
becomes piecewise constant functions $\s_i(t) \in \{-1,1 \}$ 
of an imaginary time parameter $t$, the discrete coordinate 
$\alpha \in [1,\Ns]$ being mapped to $t\in[0,\beta]$ with the correspondence 
$t = \beta \alpha/ \Ns$. In this limit the sum over $\bs$ in the
expression (\ref{eq_def_mu_ubs}) of the partition function is naturally 
interpreted as a path-integral. The discreteness of the spin degrees of 
freedom actually make such a path-integral representation~\cite{Farhi_Gutmann} 
easier to formulate than Feynman path-integrals for continuous 
coordinates~\cite{Feynman53}, and can be given a rigorous mathematical 
content~\cite{Ginibre,Gallavotti,Aiz,Ioffe}.
Note that these continuous time trajectories can be easily represented in 
the memory of a computer, as the trajectory of site $i$ 
is fully specified by $\s_i(t=0)$ and the times at which the spin flips.
Actually numerous continuous time quantum Monte Carlo algorithms do exist,
see for instance~\cite{Beard_Wiese,worm,qcluster,evertz}.

The rest of the paper will crucially rely on the
procedure developed in Sec.~\ref{sec_ct_h_cst} and \ref{sec_ct_h}. 
Though it will also be useful for
analytical purposes, it is more intuitively motivated by the following
simulational consideration. Maybe the simplest way to ensure the detailed 
balance condition in a Monte Carlo simulation which aims at sampling an 
arbitrary measure $\mu(\us)$ is to perform transitions from the current 
configuration $\us$ to a configuration obtained by replacing the value of
a randomly chosen degree of freedom $\s_i$, by a random value drawn from the 
measure conditioned on all other degrees of freedom. This procedure is known
in classical simulations as the heat-bath, or Glauber algorithm. Its
equivalent in quantum simulations consists in drawing a new configuration of
the ring $\bs_i$, or of the trajectory $\s_i(t)$ in the continuous imaginary 
time, according to the equilibrium measure induced by the spin trajectories
of all other sites. A moment of thought reveals that this boils down to study
the evolution of a single spin-1/2 in the presence of a constant transverse 
field and a piecewise constant longitudinal field, the latter being the 
effective field induced by the rest of the system on $\bs_i$. This is 
precisely the issue we shall tackle in Sec.~\ref{sec_ct_h_cst} and 
\ref{sec_ct_h}, after having recalled in Sec.~\ref{sec_ct_rev} the
well-established path-integral representation of a spin-1/2.

\subsubsection{The path integral representation of a single spin in constant
fields}
\label{sec_ct_rev}

Let us define the propagator for the evolution during an interval of imaginary
time $\lambda$ of a spin in constant transverse and longitudinal fields ($B$
and $h$ respectively):
\begin{equation}
W(\s \to \s',h,\lambda) = \langle \s | e^{\lambda (h \s^z + B \s^x ) }| \s'
\rangle \ .
\label{eq_propagator}
\end{equation}
The diagonalization of the order $2$ matrix $h \s^z + B \s^x$ easily leads to
\begin{widetext}
\begin{equation}
W(\s \to \s',h,\lambda) = 
\begin{cases} 
\cosh(\lambda\sqrt{B^2+h^2}) + \s \frac{h}{\sqrt{B^2+h^2}} 
\sinh(\lambda\sqrt{B^2+h^2}) & \text{if} \ \s=\s' \\
\frac{B}{\sqrt{B^2+h^2}} \sinh(\lambda\sqrt{B^2+h^2}) & \text{if} \ \s=-\s'
\end{cases} \ .
\label{eq_propagator_res}
\end{equation}
The path-integral representation of this propagator 
reads~\cite{Farhi_Gutmann,qcluster}
\begin{eqnarray}
W(\s \to \s, h, \lambda) &=& \sum_{n=0}^\infty B^{2n} \int_0^\lambda \de t_1 
\int_{t_1}^\lambda \de t_2 \dots \int_{t_{2n-1}}^\lambda \de t_{2 n}
\exp[\s h (2 t_1 - 2 t_2 + \dots - 2 t_{2n} + \lambda ) ] \ , 
\label{eq_pi_1} \\
W(\s \to -\s, h, \lambda) &=& \sum_{n=0}^\infty B^{2n+1} 
\int_0^\lambda \de t_1 
\int_{t_1}^\lambda \de t_2 \dots \int_{t_{2n}}^\lambda \de t_{2 n +1}
\exp[\s h (2 t_1 - 2 t_2 + \dots + 2 t_{2n+1} - \lambda ) ] \ .
\label{eq_pi_2}
\end{eqnarray}
\end{widetext}
Each term of these expressions corresponds to a spin trajectory that changes
value at times $t_1<t_2<\dots$; it is weighted by a factor $B$ raised to the 
number of such discontinuities, and by $\exp[h \int_0^\lambda \s(t)]$; 
a spin trajectory with identical (resp. opposite) initial and final value
has to jump an even (resp. odd) number of times. There are two ways
to convince oneself of the correctness of this result. Applying
the Suzuki-Trotter formalism to this single spin problem leads to such a weight
in the $\Ns \to \infty$ limit\footnote{The weight $w(\bs)$ becomes 
$\lambda B/\Ns$ raised to the number of discontinuities in the spin trajectory,
and the factors $\lambda/\Ns$ are absorbed in the change of variables from
discrete to continuous time.}. Alternatively one can notice that 
Eqs.~(\ref{eq_pi_1}),(\ref{eq_pi_2}) coincide with (\ref{eq_propagator_res})
at $\lambda =0$ and that they obey the same set of first order linear differential
equations,
\begin{eqnarray}
\frac{\partial \phantom{\lambda}}{\partial \lambda} W(\s \to \s', h, \lambda)
&=& \s' h \ W(\s \to \s', h, \lambda)
 + B \ W(\s \to - \s', h, \lambda) \ ,
\end{eqnarray}
which implies that they coincide for all values of $\lambda$.

\subsubsection{Generating trajectories for a constant longitudinal 
field}
\label{sec_ct_h_cst}

The above expressions (\ref{eq_pi_1}),(\ref{eq_pi_2}) can be interpreted as
the normalizing constants of probability measures on the set of piecewise
constant functions from $t\in[0,\lambda]$ to $\{-1,+1\}$, conditioned on their initial
($\s(t=0)$) and final ($\s(t=\lambda)$) values. More explicitly, for instance
for $\s(t=0)=\s(t=\lambda)=\s$, the probability of a trajectory with $2n$ flips
at times in the infinitesimal intervals $[t_j,t_j + \de t_j]$, with 
$t_1<\dots<t_{2n}$, is defined to be
\begin{equation}
\frac{1}{W(\s \to \s, h, \lambda)} 
B^{2n}
e^{\s h (2 t_1 - 2 t_2 + \dots - 2 t_{2n} + \lambda ) }
\, \de t_1 \dots \de t_{2n} \ .
\end{equation}
Our goal is now to construct a procedure for actually sampling from these
probability measures, that is constructing spin trajectories according to 
these weights. We shall do this by exploiting the following two identities:
\begin{widetext}
\begin{eqnarray}
W(\s \to \s,h,\lambda) &=& e^{\s h \lambda} + B 
\int_0^\lambda \de u \ e^{\s h u} \ W(-\s \to \s,h,\lambda-u) \ , 
\label{eq_recurs_W1} \\
W(\s \to -\s,h,\lambda) &=& 
B \int_0^\lambda \de u \ e^{\s h u} \ W(-\s \to -\s,h,\lambda-u) \ .
\label{eq_recurs_W2}
\end{eqnarray}
\end{widetext}
\begin{figure}
\includegraphics[width=9cm]{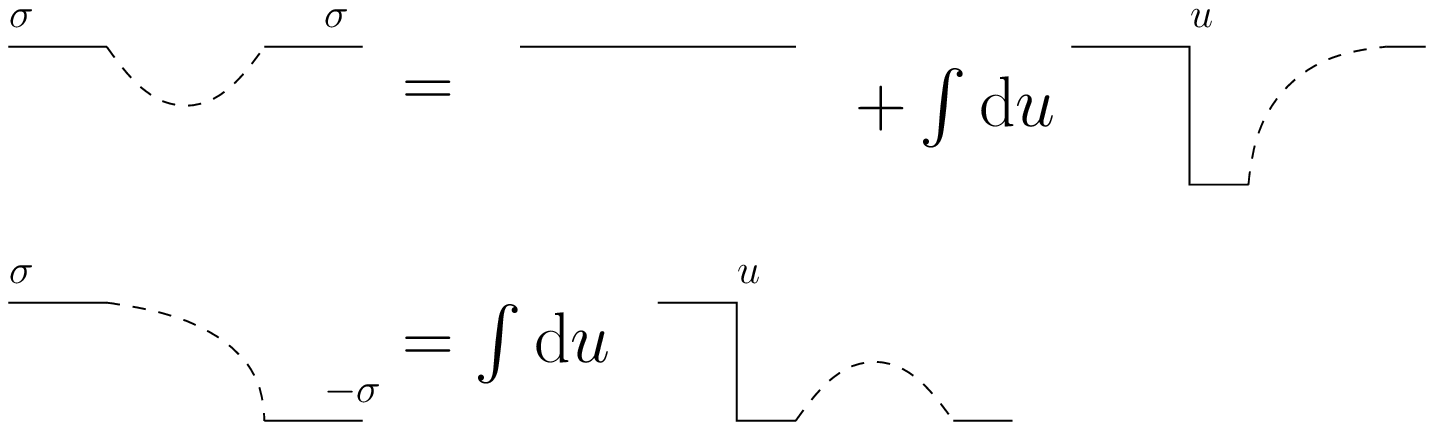}
\caption{A pictorial representation of 
Eqs.~(\ref{eq_recurs_W1}),(\ref{eq_recurs_W2}).}
\label{fig_recurs_W}
\end{figure}
The path-integral interpretation of these relations, more easily conveyed
by the drawing of Fig.~\ref{fig_recurs_W}, is as follows. 
For the first one, it means that a spin trajectory starting and ending at the 
same value $\s(0)=\s(\lambda)=\s$ is either constant on the whole time 
interval, or made of a constant part upto time $u$, followed by a jump
to $-\s$ and a second part of the trajectory representative of 
$W(-\s \to \s,h,\lambda-u)$. Similarly the second one expresses the 
necessity for a trajectory from $\s$ to $-\s$ to have at least one 
discontinuity at a given time $u$, followed by a trajectory accounting
for $W(-\s \to -\s,h,\lambda-u)$. These equalities can be proven either
from the path-integral representation of (\ref{eq_pi_1}),(\ref{eq_pi_2}), or
from the explicit expressions of $W$ given in Eq.~(\ref{eq_propagator_res}).
As a consequence of (\ref{eq_recurs_W1}),(\ref{eq_recurs_W2}) one obtains
the following recursive procedure 
to draw a spin trajectory for a constant longitudinal field on a time 
interval of length $\lambda$, constrained to $\s(0)=\s$, $\s(\lambda)=\s'$ :
\begin{itemize}
\item[$\bullet$] if $\s=-\s'$ 
\begin{itemize}
\item draw a random variable $u\in[0,\lambda]$ with density
proportional to $e^{\s h u} \ W(-\s \to -\s,h,\lambda-u)$ (see
below in eq.(\ref{gofu}) for some details on how to perform this step) 
\item set $\s(t) = \s$ upto time $u$
\item call the same procedure to generate a trajectory from $-\s$ to $-\s$
on the remaining interval of length $\lambda-u$
\end{itemize}
\item[$\bullet$] if $\s=\s'$ 
\begin{itemize}
\item with probability $e^{\s h \lambda}/W(\s \to \s,h,\lambda)$, set 
$\s(t) = \s$ on the whole time interval, and exit the procedure

\item otherwise,
\begin{itemize}
\item draw a random variable $u\in[0,\lambda]$ with density
proportional to $e^{\s h u} \ W(-\s \to \s,h,\lambda-u)$
\item set $\s(t) = \s$ upto time $u$
\item call the same procedure to generate a trajectory from $-\s$ to $\s$
on the remaining interval of length $\lambda-u$
\end{itemize}
\end{itemize}
\end{itemize}
In order to draw $u\in[0,\lambda]$ with a density 
proportional to $e^{\s h u} \ W(-\s \to -\s,h,\lambda-u)$, we compute
its cumulative distribution,
\begin{eqnarray}
G(u) &=& \frac{\int_0^u \de t \ e^{\s h t} \ W(-\s \to -\s,h,\lambda-t)}
{\int_0^\lambda \de t \ e^{\s h t} \ W(-\s \to -\s,h,\lambda-t)} =
1 - e^{\s h u}
\frac{\sinh((\lambda -u) \sqrt{B^2+h^2} )}{\sinh(\lambda \sqrt{B^2+h^2} )} \ .
\label{gofu}
\end{eqnarray}
A simple way to draw $u$ amounts to draw $G$ uniformly at random on $[0,1]$,
and to invert the above expression to obtain $u(G)$. One can proceed similarly
for the generation with the density proportional to $e^{\s h u} \ W(-\s \to
\s,h,\lambda-u)$, which involves another cumulative distribution $G$.

\subsubsection{Generating trajectories for a piecewise constant longitudinal 
field}
\label{sec_ct_h}

\begin{figure}
\includegraphics[width=9cm]{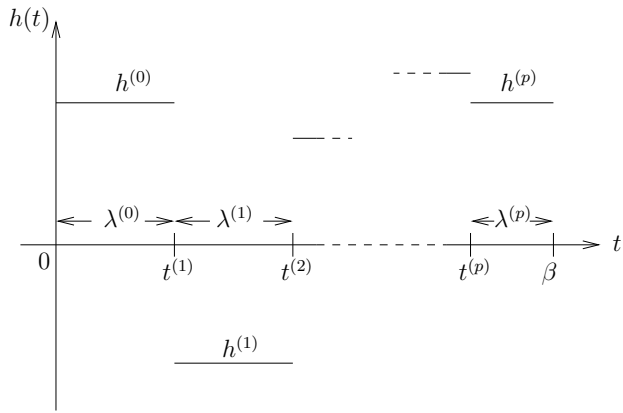}
\caption{Definition of the effective field trajectory.}
\label{fig_traj_h}
\end{figure}

In the previous subsection we considered the particular case of a constant
longitudinal field $h$. Let us now address the general case of a piecewise
constant $\bh=h(t)$ on the interval of imaginary time $[0,\beta]$, with the
following definitions illustrated in Fig.~\ref{fig_traj_h}: we shall call $p$
the number of times it changes value between $t=0$ and $t=\beta$, $0=t^{(0)}
\le t^{(1)} \le \dots \le t^{(p)} \le t^{(p+1)} = \beta$ the times of these
changes, $\lambda^{(i)} = t^{(i+1)}-t^{(i)}$ the length of these intervals for
$i\in [0,p]$, and finally $h^{(i)}$ the values the field takes in each of
these intervals.
We shall have to compute the partition function of a spin acted on by such a 
field,
\begin{equation}
\Z(\bh) = \Tr \left( 
\prod_{i=0}^p e^{\lambda^{(i)} (h^{(i)} \s^z + B \s^x )} \right) \ ,
\label{eq_def_Z_bh}
\end{equation}
and to generate spin trajectories according to the corresponding weights.
The computation of the partition function can be performed by inserting
$p+1$ representations of the identity in (\ref{eq_def_Z_bh}), corresponding
to the spin values at the imaginary times $t^{(i)}$ where $h(t)$ is
discontinuous:
\begin{eqnarray}
\Z(\bh) &=& \sum_{\s_0,\dots,\s_p} 
\Z(\s_0,\dots,\s_p | \bh )
\ , \nonumber \\
\Z(\s_0,\dots,\s_p | \bh ) &=& \prod_{i=0}^p 
W(\s_i \to \s_{i+1},h^{(i)},\lambda^{(i)}) \ ,
\label{eq_res_Zh}
\end{eqnarray}
with $\s_{p+1}=\s_0$.  Given the trajectory $\bh$ this computation is easily
performed, necessiting only the multiplication of the $p$ matrices of order
$2$ defined in (\ref{eq_propagator}). The sampling of the spin trajectory
$\s(t)$ on the interval $[0,\beta]$ is done as follows. The values
$(\s_0,\dots,\s_p)$ of the spin at times $(t^{(0)},\dots,t^{(p)})$ are
generated\footnote{This can be done with a number of operations of order
  $p$. Let us define $\Z_i(\s_0,\dots,\s_i | \bh) =
  \sum_{\s_{i+1},\dots,\s_p}\Z(\s_0,\dots,\s_p | \bh )$.  First draw
  $\s_0$ with probability $\Z_0(\s_0 | \bh)/\Z(\bh)$, then $\s_1$ according to
  $\Z_1(\s_0,\s_1|\bh)/\Z_0(\s_0 | \bh)$, and so on and so forth.} according
to the probability $\Z(\s_0,\dots,\s_p | \bh )/\Z(\bh)$.  Then for each
interval $i\in[0,p]$ a spin trajectory from $\s_i$ to $\s_{i+1}$ is generated
according to the procedure of Sec.~\ref{sec_ct_h_cst}, the longitudinal field
being constantly equal to $h^{(i)}$ on this interval of time. Finally the
$p+1$ trajectories are concatenated to obtain the full trajectory from $t=0$
to $t=\beta$.

Let us emphasize that the path integral representation of the imaginary
time evolution is well known in the 
literature~\cite{Feynman53,Farhi_Gutmann,Ginibre,Gallavotti,Aiz,Ioffe};
however we could not find in previous works such an explicit sampling 
procedure for generating the spin trajectories. Actually, as far as we know,
all continuous time quantum Monte Carlo 
algorithms~\cite{Beard_Wiese,worm,qcluster,evertz} do not proceed in an
heat-bath way, generating ``from scratch'' a new spin trajectory conditioned
on the local effective field, but rather constructing the spin update using
the current configuration itself.

\section{The quantum Bethe lattice ferromagnet}
\label{sec_qbl}

The remainder of this article will be devoted to the study of the simplest of
the finitely connected models that can be handled by the quantum cavity
method, namely the 
transverse field quantum spin-1/2 ferromagnet 
on the Bethe lattice (more precisely on a
random regular graph). The physical properties of such a model are very
intuitive: at low temperature and transverse field the model is
ferromagnetically ordered, with a positive spontaneous longitudinal
magnetization. Thermal (increasing $T$) or quantum (increasing $B$)
fluctuations destroy this order outside a region delimitated by a critical
line in the $(B,T)$ plane, that ends up in a quantum critical point at zero
temperature.

An even simpler model displaying these features is the (fully connected)
quantum Curie-Weiss model, whose solution we recall in Appendix~\ref{app_qcw}.
Both of them are of a mean-field nature, and should share most of their
qualitative properties, yet the Bethe lattice model is quantitatively
different, and technically more involved because of its finite connectivity.

\subsection{The quantum cavity method treatment}

\begin{figure}
\includegraphics[width=8cm]{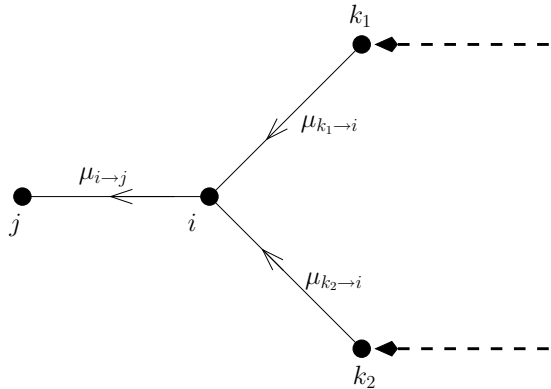}
\caption{A pictorial representation of Eq.~(\ref{eq_ferro_mu}).}
\label{fig_eq_msg_ferro}
\end{figure}

The quantum Bethe lattice ferromagnet is defined by the Hamiltonian
\begin{equation}\label{eq_H_bethe_ferro}
\hH = - \sum_{i-j} \s_i^z \s_j^z - B \sum_{i=1}^N \s_i^x \ ,
\end{equation}
where the first sum runs over the edges of a random $l+1$-regular graph of
$N$ vertices~\cite{rgraphs}. 
This means that the graph of interactions is uniformly chosen
among all graphs on $N$ vertices for which each vertex has the same number 
$l+1$ of neighbors. These graphs have the good properties to realize finite
size Bethe lattices: the set of vertices at distance\footnote{This is the 
graph theoretic distance between two vertices $i$ and $j$, defined as the 
length of a shortest path of adjacent vertices joining $i$ and $j$ along the 
graph.} smaller than a given cutoff $d$ from an arbitrarily chosen vertex 
is, with a probability which goes to one when $N$ diverges with $d$ held 
fixed, a regular tree of connectivity $l+1$. On the other hand such a graph
has no surface, contrarily to the usual Cayley tree, and the
problem of the boundary conditions for frustrated models (that will be 
encompassed by the generic treatment of Sec.~\ref{sec_qcav_generic}) is absent
with such a definition.

The hypotheses of the cavity method are more simply explained assuming first
that the graph of interactions is actually a finite tree, and that the quantum
aspects of the problem have been handled by a finite number $\Ns$ of 
Suzuki-Trotter slices. In such a case it is easy to solve the model exactly
by taking benefit of the natural recursive structure of a tree: one breaks
the graph of interaction into subtrees that are then glued together. Let
us explain this with more precise formulae, defining 
$\mu_{i\to j}(\bs_i)$ as the probability law of the configuration of the ring
$\bs_i$ when the interaction with its neighbor $j$ has been removed from the 
graph. If we denote by $\partial i$ the set of vertices neighbors of $i$, 
the recurrence equations for these distributions are:
\begin{equation}\label{eq_ferro_mu}
\mu_{i\to j}(\bs_i) = \frac{1}{z_{i\to j}} w(\bs_i) 
\prod_{k\in \partial i \setminus j } \sum_{\bs_k}
 \mu_{k\to i}(\bs_k) \, e^{\beta \bs_i \cdot \bs_k} \ ,
\end{equation}
where $z_{i \to j}$ is a normalization constant, and we introduced for
two arbitrary imaginary time dependent quantities 
$\ba \cdot \bb = \sum_{\alpha=1}^\Ns a^\alpha b^\alpha / \Ns$.
A graphical representation of this equation is given in 
Fig.~\ref{fig_eq_msg_ferro} in the case $l=2$. The site $i$ has then two 
neighbors, $k_1$ and $k_2$; the distributions $\mu_{k_{1,2} \to i}$
encode the effect on $k_{1,2}$ of the part of the tree that does not involve 
$i$; this is represented by the dashed line in figure~\ref{fig_eq_msg_ferro}. 
In absence of site $i$ the sites $k_1$ and $k_2$ are decoupled (as we
assume the graph to be a tree there are no paths between $k_1$ and $k_2$ that
do not go across $i$) and hence independent. Therefore the distribution of 
$\bs_i$ in absence of site $j$ is given by (\ref{eq_ferro_mu}). 
A proof of such recursion equations and a discussion of the connections 
with the Bethe-Peierls approximation and the Belief Propagation algorithm
can be found in~\cite{fgraph,Yedidia}.
On a given tree this
set of equations (one for each directed edge of the graph) has a unique 
solution, that is a set of ``messages'' $\mu_{i\to j}$, that can be 
efficiently determined by sweeping the edges from the leaves towards the 
center of the tree. From them all the relevant thermodynamic quantities
can be computed. For instance the probability law of the configuration of the 
ring $\bs_i$ in the complete tree is given in terms of the messages sent 
by its neighbors,
\begin{equation}
\mu(\bs_i) = \frac{1}{z_i} w(\bs_i) \prod_{k \in \partial i} 
\sum_{\bs_k} \mu_{k \to i}(\bs_k) e^{\beta \bs_i \cdot \bs_k} \ ,
\end{equation}
with $z_i$ a normalization constant, while the joint law for two neighboring 
sites $i$ and $j$ reads
\begin{equation}
\mu(\bs_i,\bs_j) = \frac{1}{z_{i-j}} \mu_{i \to j}(\bs_i)\mu_{j \to i}(\bs_j) 
e^{\beta \bs_i \cdot \bs_j} \ .
\end{equation}
Moreover the free-energy per site $f$ can be expressed in terms of the
normalization constants of these laws,
\begin{equation}\label{eq_f_cavity_singlegraph}
-\beta f = \frac{1}{N} \ln Z = \frac{1}{N}\sum_{i=1}^N \ln z_i
- \frac{1}{N} \sum_{i-j} \ln z_{i-j} \ ,
\end{equation}
where the second sum runs over the (undirected) edges of the tree. 

The above derivation was exact because we assumed the graph of interactions
to be a tree. The scope of the cavity method is to extend these results to
random graphs which are only locally tree-like, in the precise sense explained
above. In its simplest version, called replica symmetric for historical 
reasons, one assumes the existence of a single pure state in the configuration
space of the random graph model. This implies a decay of correlations at large
distance in the graph, hence the effect of the long loops neglected in the tree
derivation amounts to create a self-consistent boundary condition which
traduces the absence of a surface in the random graph. As in the present model
all sites have the same neighborhood (there is no fluctuation neither in the 
connectivity nor in the intensity of the interaction couplings), one has to
look for a solution of (\ref{eq_ferro_mu}) where the messages 
$\mu_{i\to j}$ are all equal to a single law, that we shall denote in the
following $\eta(\bs)$, which is seen from (\ref{eq_ferro_mu}) to satisfy
\begin{equation}
\eta(\bs) = \frac{1}{z_l} w(\bs) 
\left(\sum_{\bs'} \eta(\bs') e^{\beta \bs \cdot \bs'}  \right)^l \ .
\label{eq_ferro_eta}
\end{equation}
The assumption on the unicity of the pure state can fail for two kind of 
reasons. In ferromagnetic models, as the one considered now, there is an 
ordered phase at low temperature and transverse field in which two pure 
states coexist because of the up/down symmetry of the model. This is not a 
serious limitation of the method, in the following it will be kept understood
that an infinitesimal longitudinal field is applied to the system in order to
select one of the two pure states. In fact the exactness of the cavity method
for classical ferromagnetic models on random graphs has been recently proven
rigorously~\cite{ferro_rigorous}. A much more serious problem, that shall not
be discussed in this paper, arises in frustrated models, when an exponential 
number of pure states proliferate at low temperatures; the replica-symmetry
breaking version of the cavity method is then required to solve the problem.

Let us first write the solution of the equation (\ref{eq_ferro_eta}) 
in the classical situation,
for $B=0$. In such a case the weight $w$ forces all spins along the ring to
take the same value, hence
\begin{eqnarray}
\eta(\bs) &=& \frac{1+\tanh (\beta h)}{2} 
\left( \prod_{\alpha=1}^\Ns \delta_{\s^\alpha,1} \right) +
\frac{1-\tanh (\beta h)}{2} 
\left(\prod_{\alpha=1}^\Ns \delta_{\s^\alpha,-1} \right) \ ,
\end{eqnarray}
where $h$ is solution of
\begin{equation}
h=\frac{l}{\beta} \text{arctanh} 
\left(\tanh(\beta)  \tanh(\beta h)  \right) \ .
\label{eq_h_class}
\end{equation}
The classical model thus exhibits a continuous paramagnetic to ferromagnetic
transition when the inverse temperature $\beta$ crosses its critical value, 
$\beta_{\rm c} = \text{arctanh} (1/l)$.

We now come back to the general $B$ case and rewrite explictly the cavity
method prediction for the free-energy per site, that is obtained from
Eq.~(\ref{eq_f_cavity_singlegraph}) by substituting $\mu_{i\to j} = \eta$:
\begin{eqnarray}
-\beta f &=& \ln \left( 
\sum_{\bs} w(\bs) \left( 
\sum_{\bs'} \eta(\bs') e^{\beta \bs \cdot \bs'}\right)^{l+1}   \right) 
- \frac{l+1}{2} \ln\left( 
\sum_{\bs, \bs'} \eta(\bs) \eta(\bs') e^{\beta \bs \cdot \bs'}  \right) \ ,
\label{eq_qbl_f_var}
\end{eqnarray}
and for the probability law on one site and two adjacent sites,
\begin{eqnarray}
\mu(\bs) &=& \frac{1}{z_{l+1}} w(\bs) 
\left(\sum_{\bs'} \eta(\bs') e^{\beta \bs \cdot \bs'}  \right)^{l+1} \ ,
\nonumber \\
\mu(\bs,\bs') &=& \frac{1}{z_{\rm edge}} 
\eta(\bs) \eta(\bs') e^{\beta \bs \cdot \bs'} \ . 
\label{eq_marginal_qbl}
\end{eqnarray}
Note that the expression (\ref{eq_qbl_f_var}) is variational, in
the sense that its derivative with respect to $\eta$
vanishes on the solution of equation (\ref{eq_ferro_eta}).

\begin{figure}
\includegraphics[width=8cm]{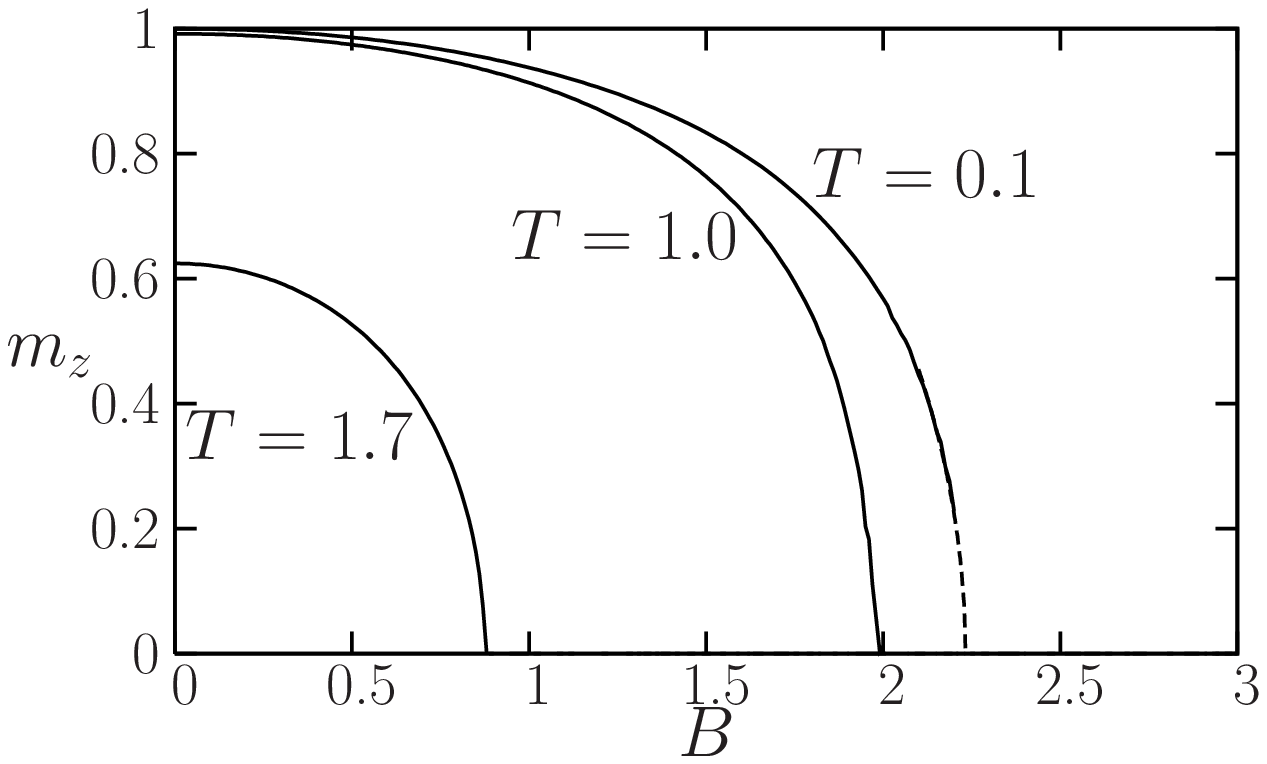}
\includegraphics[width=8cm]{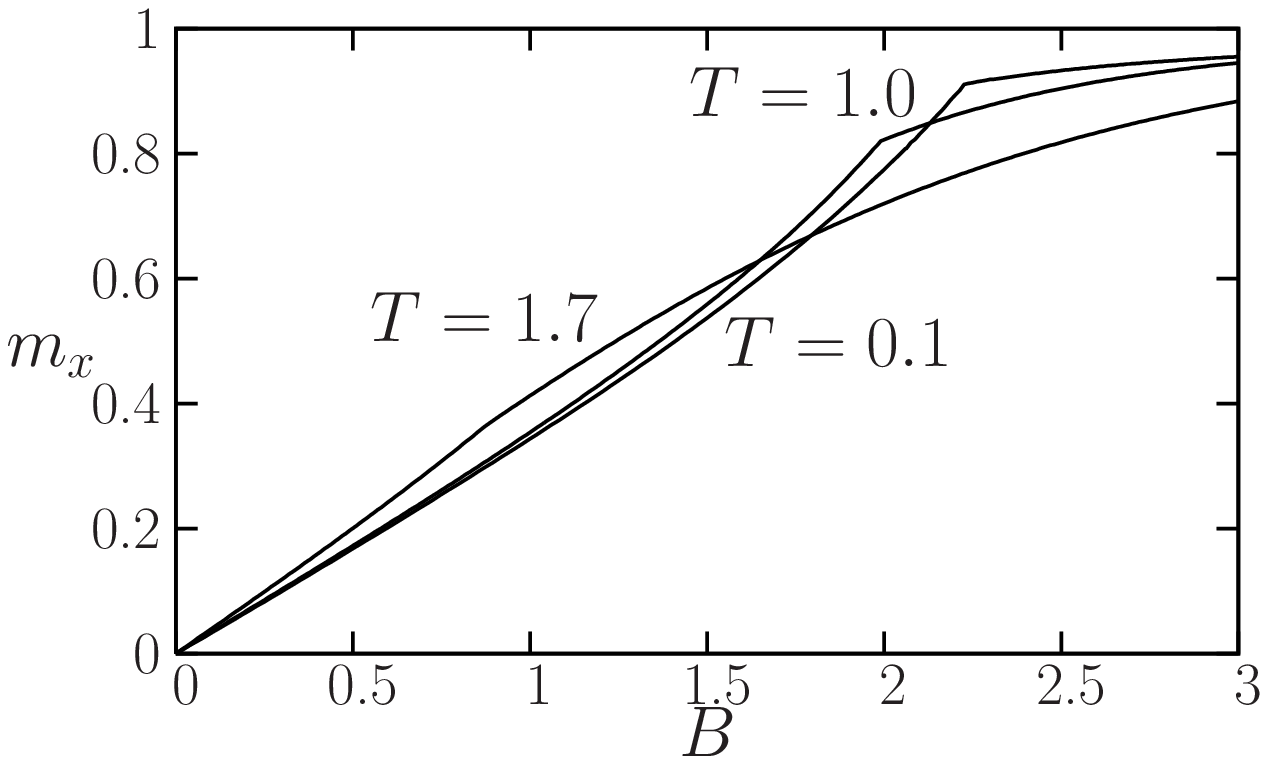}
\caption{Longitudinal ($m_z$, left panel) and transverse ($m_x$, right panel) 
magnetizations as functions of the transverse
field $B$ for $T=1.7,1.0,0.1$. For the lowest temperature we could not obtain reliable data for $m_z$
close to the critical field due to critical fluctuations. The dashed line is a fit to 
$m_z \sim (B_c-B)^{1/2}$ done in the interval $B\in [2.1,2.2]$ and serves as a guide to the eye.
}
\label{fig_mxz_continuous}
\end{figure}

\subsection{A convenient representation of $\eta(\bs)$}
\label{sec_qbl_rep}

We turn now to the problem of the determination of the probability law
$\eta(\bs)$ solution of Eq.~(\ref{eq_ferro_eta}). As this is a law on the
probability of $\Ns$ Ising spins, its complete characterization should involve
$2^\Ns - 1$ real numbers. Such a direct representation becomes very soon
impossible to handle when $\Ns$ grows, and in particular in the continuous time
limit $\Ns \to \infty$. We shall however see that an alternative
representation allows to bypass this difficulty.
First we define a probability law $p(\bs|\bh)$ on the configurations of
a ring of spins $\bs$ by
\begin{equation}
p(\bs|\bh) = \frac{1}{\Z(\bh)} 
\ w(\bs) \  e^{\beta \bs \cdot \bh} \ , \qquad 
\Z(\bh) = \sum_\bs w(\bs) \  e^{\beta \bs \cdot \bh} \ ,
\label{eq_def_p}
\end{equation}
$\Z(\bh)$ ensuring the normalization of $p(\bs|\bh)$.
These definitions allow to rewrite Eq.~(\ref{eq_ferro_eta}) as
\begin{widetext}
\begin{equation}
\label{eq_eta_popu_ferro}
\eta(\bs) = \sum_{\bs_1,\dots,\bs_l} \eta(\bs_1) \dots \eta(\bs_l)
\ p(\bs|\bs_1 + \cdots + \bs_l) \ \frac{\Z(\bs_1 + \cdots + \bs_l)}{z_l} \ .
\end{equation}
\end{widetext}
Suppose now that one has an estimation of $\eta(\bs)$ given by a
representative weighted sample of $\Nt$ elements $\{ \bs_i \}$, that is
\begin{equation}\label{eq_eta_popu}
\eta(\bs) = \sum_{i=1}^\Nt a_i \ \delta(\bs - \bs_i ) \ ,
\end{equation}
such that the weights $a_i$ add up to one. A new estimation of $\eta$
(i.e. a new set of configurations $\bs_i'$ and weights $a_i'$) can
then be obtained by plugging this estimation in the left-hand-side of
Eq.~(\ref{eq_eta_popu_ferro}), which leads to the following procedure.
To generate each of the new $\Nt$ representants of $\eta$, one repeats in an
independent way the steps:
\begin{itemize}
\item[$\bullet$] draw independently $l$ integers $i_1,\dots,i_l$ in $[1,\Nt]$
with probability $a_i$
\item[$\bullet$] set $\bh=\bs_{i_1}+\dots+\bs_{i_l}$
\item[$\bullet$] generate a configuration $\bs_i'$ according to the law
$p(\bs | \bh)$, and set $a_i' = \Z(h)$
\end{itemize}
Once the $\Nt$ new elements have been generated one just has to multiply the
weights $a_i'$ by a global normalization factor, and the new estimates can be
again plugged in the right-hand-side of Eq.~(\ref{eq_eta_popu_ferro}) to
approach its fixed point solution.

At this point the continuous time limit $\Ns \to \infty$ can be taken without
any difficulty of principle: as long as $\beta$ is finite the spin
trajectories have only a finite number of discontinuities, and can then be
easily encoded with a finite set of reals corresponding to the time they
change values. Moreover one can easily show that in this limit drawing a spin
trajectory from the law $p(\bs|\bh)$ corresponds exactly to the procedure we
defined in Sec.~\ref{sec_ct} (we show explicitely this correspondence for the
normalization $\Z(\bh)$ in App.~\ref{app_details_ct}).

This representation of the probability law $\eta(\bs)$ by a sample of
representative elements is a widespread technique in the field of disordered
systems~\cite{popu_first,MePa}. We warn however the reader accustomed to the
classical cavity method that we did not use it exactly in the usual way, as
the classical equivalent of $\eta(\bs)$ is a single real number, not a
probability law.

Before closing this section let us discuss the computation of the physical
observables in the continuous limit, taking as representative examples the
longitudinal and transverse magnetizations. These can be obtained by taking
the $\Ns \to \infty$ limit of the formulas 
(\ref{eq_obs_diag}),(\ref{eq_mx_nbfinite}), which yields
\begin{eqnarray}
m_z &=& \lim_{N \to \infty} \frac{1}{N} \sum_{i=1}^N \langle \s_i^z \rangle = 
\sum_\bs \mu(\bs) \frac{1}{\beta} \int_0^\beta \de t \ \s(t) \ , \nonumber \\
m_x &=& \lim_{N \to \infty} \frac{1}{N} \sum_{i=1}^N \langle \s_i^x \rangle = 
\sum_\bs \mu(\bs) \frac{1}{\beta B} j(\bs) \ ,
\end{eqnarray}
where in the right-hand-side $\mu(\bs)$ is to be computed from
Eq.~(\ref{eq_marginal_qbl}), the sums over $\bs$ are understood as sums
over continuous time trajectories, and we have defined $j(\bs)$ as the
number of discontinuities of $\s(t)$ on the interval $[0,\beta]$.

\begin{figure}
\includegraphics[width=8cm]{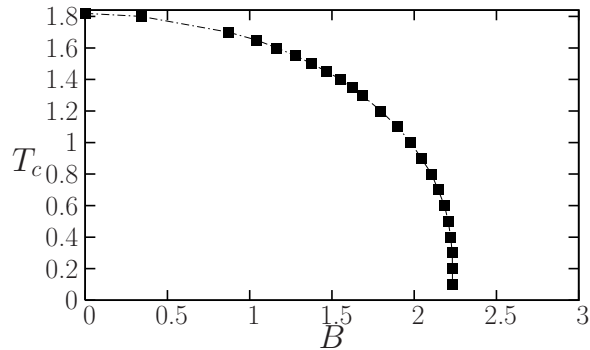}
\caption{
Phase diagram: the critical temperature $T_{\rm c}$ as a function of the transverse field $B$.
}
\label{fig_phase_diag_continuous}
\end{figure}

\subsection{Continuous-time results}

We present now numerical results for $l=2$; the behavior is qualitatively the
same for any $l$. We followed the method of resolution presented above,
using $\Nt=10^4$ trajectories.
For a fixed temperature, we initialized the population 
at $B=0$; then each trajectory is constant and its value is chosen in such a way
that the average magnetization is equal to 
the classical magnetization that can be obtained from the solution of Eq.~(\ref{eq_h_class}).
In this way we select one of the two possible ferromagnetic states, 
that we can follow by increasing gradually the transverse field. We increased $B$
with a step $dB = 10^{-2}$ and for each step we let the population equilibrate
for $10^2$ iterations and averaged the observables over $10^3$ subsequent steps.
We checked that the weights $a_i$ defined in (\ref{eq_eta_popu}) remain quite uniformly
distributed over the population at all investigated temperatures and transverse fields.

In figure~\ref{fig_mxz_continuous} we plot the magnetizations $m_z$ and
$m_x$ as functions of the transverse field $B$
for three different temperatures 
$T < T_{\rm c}(B=0) = 1/\text{arctanh}(1/2) = 1.820\ldots$.
At these low temperatures, the system undergoes
a continuous phase transition at a critical value $B_{\rm c}(T)$ of the transverse field. The transition
is characterized by the vanishing of $m_z$ and by a jump in the derivative of $m_x$. Unfortunately
obtaining reliable values of $m_z$ close to $B_{\rm c}$ is a difficult numerical task due to critical
fluctuations that cause strong finite size effects in the size of the population ${\cal N}_{\rm traj}$.
Therefore we located the transition by the jump in the derivative of $m_x$; we fitted $m_x$ by
linear laws close to $B_{\rm c}$ from above and below, and determined their intersection. 

\begin{figure}
\includegraphics[width=8cm]{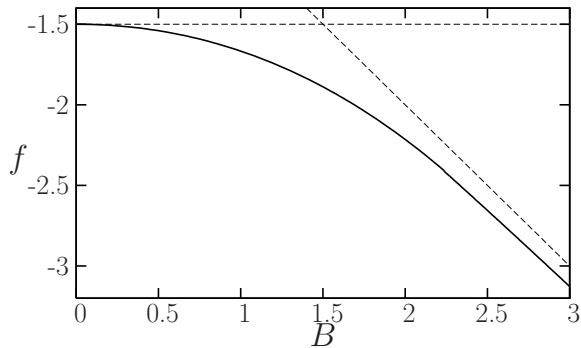}
\caption{ Free energy as a function of $B$ for $T=0.1$. The curves for
  $T=0.3,0.2,0.1$ coincide within our numerical precision. Therefore we assume
  that this curve is representative of the ground state energy as a function
  of $B$. The asymptotic values for $B\to 0$ ($e_{GS} = 3/2$) and for $B\to
  \infty$ ($e_{GS}=-B$) are reported as dashed lines.  The subleading
  correction for large $B$ is $\propto B^{-1}$ and is not reported.  }
\label{fig_ground_state_e}
\end{figure}

The resulting phase diagram is reported in fig.~\ref{fig_phase_diag_continuous}.
We found that for $T\lesssim 0.3$ the temperature dependence of all observables is very weak 
and $B_{\rm c} \sim 2.232$,
that is a reliable estimate for the $T\to 0$ limit and is in excellent agreement with the
value determined in~\cite{qbl}. 
The scaling of $B_{\rm c}$ for $T\to 0$ is compatible with the essential singularity that is
found in the Curie-Weiss model,
see Appendix~\ref{app_qcw} and in particular Eq.~(\ref{eq_BcT0_qcw}).
The very weak (practically unobservable) temperature dependence of the free energy below
$T=0.3$ make us confident that the entropic term is negligible at these low temperatures
and the free energy for $T=0.1$, plotted in figure~\ref{fig_ground_state_e} 
is representative of the ground state energy $e_{GS}$. The latter has a weak singularity
(discontinuity of the second derivative) at $B_{\rm c}$ which is not observable with our
numerical precision, but is easily observed by a direct computation of 
$m_x = -\frac{de_{GS}}{d B}$, see figure~\ref{fig_mxz_continuous}.

Overall, our results are in very good quantitative 
agreement with the ones obtained at $T=0$ in~\cite{qbl} by a matrix product 
state description, except for the value of the exponent $\beta$ characterizing
the vanishing of the magnetization close to $B_{\rm c}$, $m_z \sim (B_{\rm c}-B)^\beta$.
Even if we do not have very precise data close to $B_{\rm c}$ due to finite population-size
effects, our data are compatible with the mean-field exponent $\beta=0.5$ at all investigated
temperatures, while in~\cite{qbl} a slightly smaller value of $\beta \sim 0.41$ is reported
at $T=0$. We will further comment on this discrepancy in section~\ref{sec_FSS}.

\subsection{Comparison with approximate treatments}

In this section we compare the previous results with approximated solutions
of Eq.~(\ref{eq_ferro_eta}). First we consider the finite $\Ns$ case, then we consider
variational approximations to the solution.

\subsubsection{Resolution at finite $\Ns$}

At finite $\Ns$, the distribution $\eta(\bs)$ can be encoded with $2^\Ns-1$
real numbers (e.g. the probabilities of each of the $2^\Ns$ configuration, with
their sum constrained to be 1). Then Eq.~(\ref{eq_ferro_eta}) can be rewritten
as a fixed point equation for these numbers, and the solution can be found by
simple iteration (below we refer to this procedure as ``exact solution'' for finite $\Ns$). 
This method is extremely precise but computationally very heavy,
as the time to solve the equation scales exponentially in $\Ns$. We are then
limited to $\Ns\leq 13$; the computation for the largest $\Ns=13$ took two weeks
on a standard workstation, while we estimated the one for $\Ns=14$ to take at least
two months.

Therefore, in order to study the approach to the limit $\Ns \to \infty$, we resorted to
the population method described in Sec.~\ref{sec_qbl_rep}.
We represented $\eta(\bs)$ by a population of discrete time
trajectories $\bs = ( \s^1,\cdots,\s^\Ns )$ and solved 
Eq.~(\ref{eq_ferro_eta}) by iteration following the procedure detailed in 
section~\ref{sec_qbl_rep} without taking the limit $\Ns \to \infty$.
Similarly to the continuous time case, we used ${\cal N}_{\rm traj} = 10^5$ and 
performed $10^2$ iterations to achieve
convergence, after which data have been collected along $10^3$ iterations.
The computation time is now polynomial in $\Ns$ so we can go to much larger values
(at the price of numerical precision due to finite ${\cal N}_{\rm traj}$).
Note that the continuous time computation is much more
efficient. In the discrete time case, at low $B$ and large $T$, 
the spins $\s^\alpha$ are constant along large time intervals and the information encoded
in a discrete trajectory is redundant. 
On the contrary, at low $T$ and large $B$ the method at finite $\Ns$ is obviously 
incorrect because of the discretization. One could think to adapt $\Ns$ in order
to find the optimal value for a given $T$,$B$; however this is most naturally done
in the continuous-time framework where the number of spin flips is the natural variable
describing a trajectory. Therefore we think that the finite $\Ns$ population method is
useful only for the illustrative purposes of this section.

\begin{figure}
\includegraphics[width=8cm]{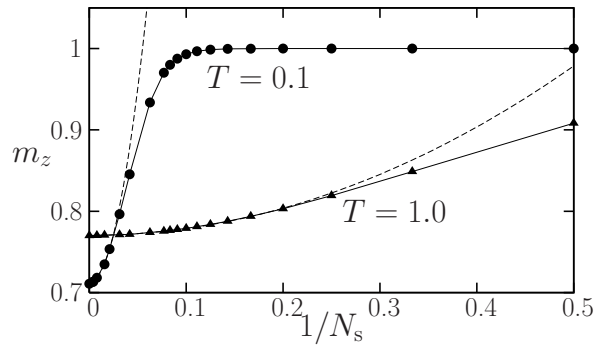}
\caption{
Longitudinal magnetization $m_z$ as a function of $1/\Ns$ for $T=1.0$, $B=1.5$ (triangles)
and for $T=0.1$, $B=1.8$ (circles). In both cases the point at $1/\Ns=0$ corresponds to the
continuous time result.
Full lines are guides to the eye, dashed lines are fit to quadratic laws, 
$m_z - m_z(\Ns \to \infty) \propto \Ns^{-2}$.
}
\label{fig_finite_Ns_mz}
\end{figure}

\begin{figure}
\includegraphics[width=8cm]{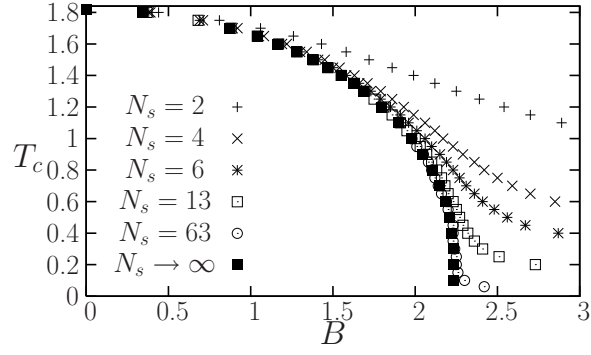}
\caption{
Phase diagram at finite $\Ns$: 
the critical temperature $T_{\rm c}$ as a function of the transverse field $B$.
Note that for a fixed $\Ns$, $T_{\rm c}(B\to \infty) = T_{\rm c}(B=0)/\Ns$.
Filled squares are the result of the continuous time method, already reported in
figure~\ref{fig_phase_diag_continuous}.
}
\label{fig_phase_diag_Ns}
\end{figure}

Data reported in this section and in figures~\ref{fig_finite_Ns_mz}, \ref{fig_phase_diag_Ns} 
have been obtained by exact solution for $\Ns \leq 13$ 
and by the population method for $\Ns > 13$.
We do not report detailed results for the magnetization as a function of $B$; as
expected, the agreement with the continuous time computation is very good 
at high temperature, and becomes poorer and poorer on lowering the temperature. 
We focused on two points in the phase diagram; the first at intermediate temperature
$T=1.0$ and $B=1.5$; the second at very low temperature $T=0.1$ and $B=1.8$. In
figure~\ref{fig_finite_Ns_mz} we plot the longitudinal magnetization $m_z$ for
different values of $\Ns$. As expected, finite $\Ns$ effects are much stronger
at low temperature. Note that for large $\Ns$ the leading correction is 
$\propto \Ns^{-2}$, as in the Curie-Weiss model (see Appendix~\ref{app_qcw}).
For $T=1$ deviations from the leading term are very small and
$\Ns \sim 10$ is enough to get a fair estimate of the true $m_z$. On the contrary,
for $T=0.1$ deviations are very large and $\Ns \gtrsim 256$ is needed.

In figure~\ref{fig_phase_diag_Ns} we report the critical temperature as a function
of $B$ for different values of $\Ns$ together with the continuous time result.
Note that in the limit
of large transverse field, each of the $\Ns$ time slices becomes independent of the others.
The model reduces to $\Ns$ copies of the classical system ($B=0$) with a ferromagnetic
coupling rescaled by $1/\Ns$. Then in this limit the critical temperature is given by
the classical one divided by $\Ns$, $T_{\rm c}(\Ns,B\rightarrow \infty) = 
1/[\Ns \text{arctanh}(1/l)]$.
Below this temperature the system is always in the ferromagnetic phase, so that
the quantum phase transition cannot be studied within this approximation.

For generic spin-$1/2$ models, the finite $\Ns$ approximation might be useful in order to
understand the behavior at intermediate temperatures, and might also
give quantitatively accurate results far from $T=0$. This might be useful
in cases where more complicated cavity treatments (such as the so-called
{\it one-step replica symmetry breaking}) are necessary. Still the continuous time
method appears to us preferable as it remains reliable down to very low temperatures.

\subsubsection{The static approximation}

A different strategy to obtain approximate solutions of these models is to use the variational
property of the Bethe free energy (\ref{eq_qbl_f_var}): the free energy of the model is 
the minimum of the free energy defined by Eq.~(\ref{eq_qbl_f_var}) over the 
function $\eta(\bs)$ (in the classical case the correctness of this procedure 
has been proven in \cite{ferro_rigorous}).
Then one can propose a variational form
for $\eta(\bs)$ and minimize the free energy with respect to the free parameters to
obtain an upper bound to the true free energy of the problem.

A popular variational form is the so-called
{\it static approximation}, that in our context amounts to the following
{\it ansatz} for $\eta(\bs)$:
\begin{equation}
\eta(\bs) = \eta\left[\frac1\beta \int_0^\beta \de t \, \sigma(t)\right]=
\eta\left[\frac1\Ns \sum_{\alpha=1}^\Ns \sigma_\alpha\right] \ ,
\end{equation}
i.e. the probability of a trajectory depends only on its average spin value 
along the imaginary time.
This approximation has been widely used
in computations based on the replica method for fully connected 
models~\cite{BrayMooreQSK,Nish_stat,Goldschmidt,Leticia} 
and has been recently applied to the random $k$-satisfiability in a transverse
field~\cite{qksat}.
The present derivation based on the cavity method gives back the same results originally
derived in \cite{qksat}, but is simpler because the use of replicas
is avoided. Here we discuss only the case of the ferromagnet on a regular graph but the
equations can be easily generalized to more complicated cases where fluctuations of the couplings
and/or the connectivity are present~\cite{qksat}.

Equivalently we can rewrite the equation above as
\begin{eqnarray}
\eta(\bs) &=& \int_{-\infty}^{\infty} \de h \, p(h) \prod_{\alpha=1}^\Ns \left[ \frac{e^{h \sigma_\alpha}}{2 \cosh h} \right]
= \int_{-1}^{1} \de m \, p(m) \prod_{\alpha=1}^\Ns \left[ \frac{1+m \sigma_\alpha}2 \right]
\ ,\label{eq_eta_var_static}
\end{eqnarray}
where with a slight abuse of notation we used $p$ for the distribution of
both the effective field $h$ and its associated magnetization $m = \tanh h$.
This expression makes evident that the assumption of the static approximation is
that the spins in a trajectory are uncorrelated along the imaginary time,
and subject to a common field extracted from the distribution $p(h)$.

From Eq.~(\ref{eq_eta_var_static}) it follows that, in the limit $\Ns \to \infty$:
\begin{widetext}
\begin{equation}\label{eq_conti_static}
\begin{split}
\sum_{\bs'} \eta(\bs') & e^{\beta \bs \cdot \bs'} = \int_{-1}^1 \de m \, p(m) 
\prod_{\alpha=1}^\Ns \left[ \cosh \frac{\beta}\Ns + m \sigma_\alpha \sinh \frac{\beta}\Ns \right] 
\sim \int_{-1}^1 \de m \, p(m) 
\prod_{\alpha=1}^\Ns e^{ m \sigma_\alpha \beta/\Ns } \ , \\
\sum_{\bs, \bs'} \eta(\bs) & \eta(\bs') e^{\beta \bs \cdot \bs'} = \int_{-1}^1 
\de m \, \de m' \, p(m) p(m') 
e^{\beta m m'} \ , \\
\sum_{\bs} w(\bs) &
\left( \sum_{\bs'} \eta(\bs') e^{\beta \bs \cdot \bs'}\right)^{l+1} =
\int_{-1}^1 \de m_1 \, p(m_1) \cdots \de m_{l+1} \, p(m_{l+1}) \sum_\bs w(\bs) \, 
e^{\beta\big[ \sum_{p=1}^{l+1} m_p\big]
\big[ \Ns^{-1}\sum_{\alpha=1}^\Ns
\sigma_\alpha \big]  } \\
&=\int_{-1}^1 \de m_1 \, p(m_1) \cdots \de m_{l+1} \, p(m_{l+1}) \int_{-1}^1 
\de m
\, e^{\beta m \sum_{p=1}^{l+1} m_p  } \sum_\bs w(\bs) \,
\delta\left(m-\frac1\Ns\sum_{\alpha=1}^\Ns \sigma_\alpha \right) \\
&=\int_{-1}^1 \de m_1 \, p(m_1) \cdots \de m_{l+1} \, p(m_{l+1}) \int_{-1}^1 
\de m
 \, e^{\beta m \sum_{p=1}^{l+1} m_p  } \, w_{\rm s}(m)  \ ,
\end{split}
\end{equation}
\end{widetext}
where in the last line we defined the function
\begin{equation}\label{eq_u0}
w_{\rm s}(m) =  \sum_\bs w(\bs) \,
\delta\left(m-\frac1\Ns\sum_{\alpha=1}^\Ns \sigma_\alpha \right) \ .
\end{equation}
The following explicit expression for $w_{\rm s}(m)$ can be found 
in~\cite{qksat} (where it was called $e^{-\beta u_0(m)}$):
\begin{equation}
w_{\rm s}(m) = \frac{\beta B}{\sqrt{1-m^2}} I_1\big(\beta B \sqrt{1-m^2}\big)
+ \delta(m-1) + \delta(m+1) \ ,
\label{eq_ws}
\end{equation}
with $I_1$ the modified Bessel function of the first kind. For completeness
we provide a proof of this result in App.~\ref{app_ws}.
From the equations above one obtains the free energy (\ref{eq_qbl_f_var})
as a functional of $p(m)$, that has then to be minimized.
Remarkably, the resulting expression for the free energy is equivalent to
the cavity free energy for a classical system whose variables are
the continuous $m_i \in [-1,1]$ and whose Gibbs measure is defined by
\begin{equation} 
\mu_{\rm s}(\underline m) = 
\frac1{Z_{\rm s}} e^{-\beta H_{\rm s}(\underline m)} = 
\frac1{Z_{\rm s}} e^{ \beta \sum_{i-j} m_i m_j} \prod_{i=1}^N w_{\rm s}(m_i) \ ,
\end{equation}
i.e. it is obtained from the quantum measure (\ref{eq_def_mu_ubs}) 
by replacing the quantum
operators by their average and the transverse field term by its average
as defined in (\ref{eq_u0}).
The analogy with a classical system, or a direct differentiation of the
free energy with respect to $p(m)$, allows to write a cavity equation
for $p(m)$ which is the analog of (\ref{eq_ferro_eta}):
\begin{equation}\label{eq_static_cavity}
p(m) = \frac{1}{z_l} w_{\rm s}(m) \left( \int_{-1}^1 \de m' \, p(m') 
e^{\beta m m'} \right)^l \ .
\end{equation}
Given the structure of $w_{\rm s}(m)$ it turns out that
$p(m)$ can be decomposed as 
$ a_+ \delta(m-1) + a_- \delta(m+1) + \widetilde p(m)$, where 
$\widetilde{p}$ has its support strictly between $-1$ and $1$.
We solved Eq.~(\ref{eq_static_cavity}) by iteration, using a discretized
representation of the regular part $\widetilde{p}$ and keeping explicitly
the weights $a_\pm$.
Note that at $B=0$ the first term in (\ref{eq_ws}) vanishes. Then $a_\pm = (1\pm \tanh(\beta h))/2$ 
where $h$ is the solution of Eq.~(\ref{eq_h_class}): the static approximation is obviously
exact in the classical case.
For a given temperature $T < T_{\rm c}(B=0)$, we initialised $p(m)$ on the classical solution
and increased gradually the transverse field $B$, at each step iterating Eq.~(\ref{eq_static_cavity})
until convergence (typically after $\sim 10^2$ iterations with a discretization step 
$dm \sim 10^{-3}$). 

\begin{figure}
\includegraphics[width=8cm]{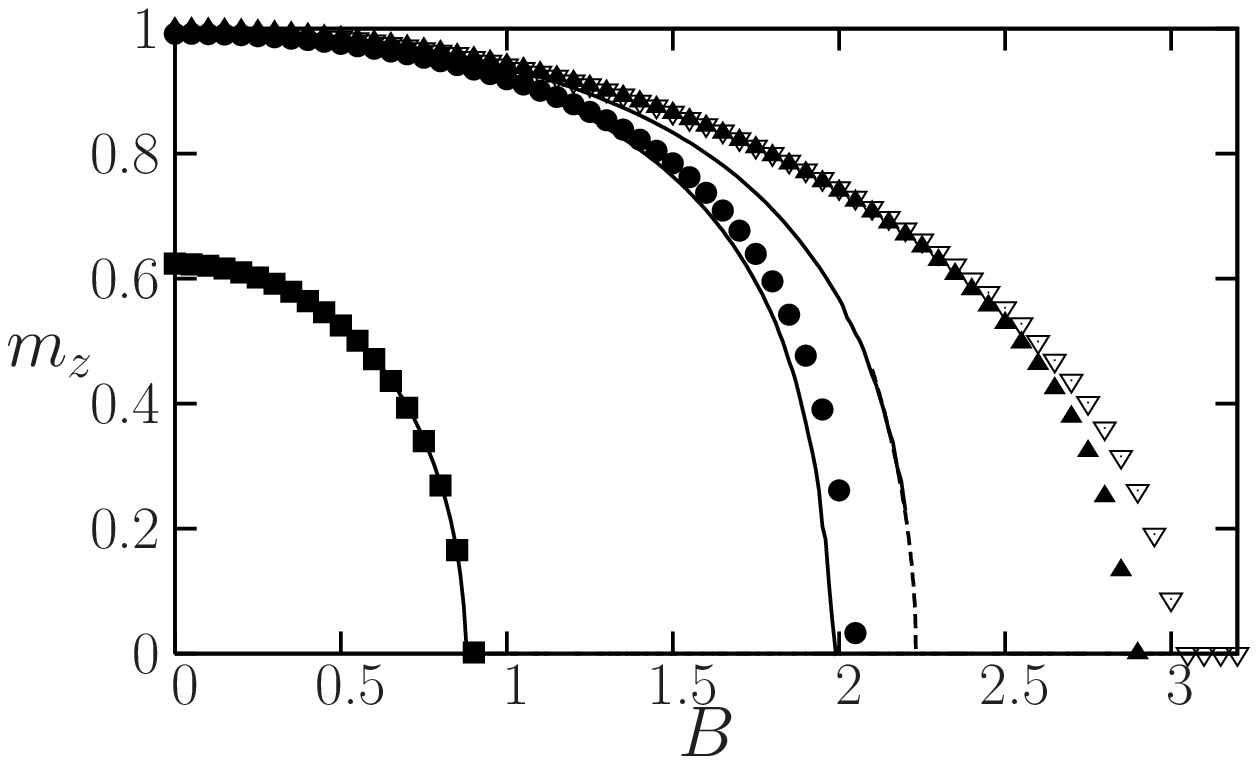}
\includegraphics[width=8cm]{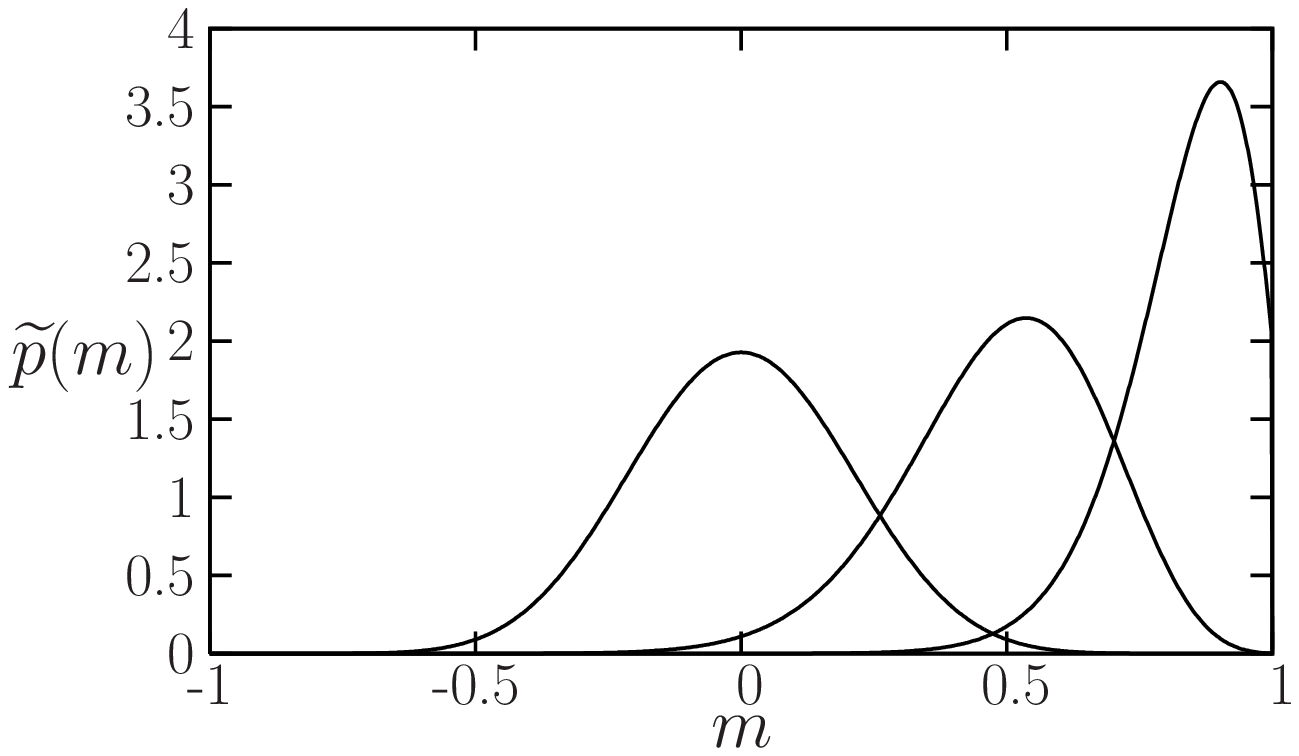}
\caption{(Left panel) Longitudinal magnetization $m_z$ as a function
of the transverse field $B$. 
The full lines are the exact results
reported in figure~\ref{fig_mxz_continuous} for $T=1.7,1.0,0.1$ 
(from left to right). Points are the results of the static 
approximation: $T=1.7$ (squares), $T=1.0$ (circles), $T=0.1$
(triangles), $T=0$ (open triangles).
(Right panel) The function $\widetilde p(m)$ defined in the text at
temperature $T=0.1$ and transverse field $B=3.2,2.2,1.2$ (from left to
right). For smaller values of $B$, the maximum approaches $m=1$ and
the weight of the $\delta(m-1)$ starts to increase until 
$\widetilde p(m)$ vanishes at $B=0$.
}
\label{fig_static}
\end{figure}

Eq.~(\ref{eq_static_cavity}) can be further simplified in the limit $T\to 0$. 
As $I_1(x) \sim e^x$ for large $x$, the Dirac distributions can be neglected
in (\ref{eq_ws}) for any $B>0$, and at the leading exponential order
$w_{\rm s}(m) \sim e^{\beta B \sqrt{1-m^2}}$. We can thus use the simplified
expressions $p(m) \sim e^{\beta \zeta(m)}$, $z_l \sim e^{-\beta e_l}$ at this
order, and hence obtain from (\ref{eq_static_cavity}):
\begin{equation}
\zeta(m) = e_l + B \sqrt{1-m^2} + l  \max_{m'} [ m m' + \zeta(m') ] \ ,
\end{equation}
where the normalization $e_l$ must be determined in such a way that 
$\max_m \zeta(m)=0$.
Like Eq.~(\ref{eq_static_cavity}), the latter equation can be solved iteratively. In this case
a good starting guess is $\zeta_0(m) =  B \sqrt{1-m^2} + h_0 m + \text{const}$, 
where the symmetry $m \to -m$ is initially 
broken by the term $h_0 m$ (with $h_0$ reasonably small) in order to select one state. 
Also in this case we used a discretization step 
$dm \sim 10^{-3}$ and observed convergence after $\sim 10^2$ iterations.

The results of the static approximation are reported in figure~\ref{fig_static}.
In the left panel we compared $m_z$ obtained from the static approximation
with the exact one obtained by the continuous time solution of the cavity
equation. As expected the approximation is very good for $T \gtrsim 1.$
and becomes poor close to $T=0$. Still, the qualitative prediction of the
static approximation remain reliable down to $T=0$, even if the value of 
$B^{\rm static}_{\rm c}(T=0) = 3.0$ predicted by the static approximation is
different from the exact one, $B_{\rm c}(T=0) = 2.232$. In the right panel
of figure~\ref{fig_static} we show the typical shape of $p(m)$: it is a symmetric
function at large $B$, where $a_\pm =0$. On lowering $B$
below $B_{\rm c}^{\rm static}$, the distribution becomes asymmetric and 
$a_+ > a_- > 0$, even if $a_\pm$ remain very small at intermediate $B$. On approaching
$B=0$, $a_\pm$ grow faster and $\widetilde p(m)$ vanishes until the classical solution
is recovered.

In summary, the static approximation is a reliable variational tool to study
qualitatively the phase diagram of the system. Remarkably, it can be solved
down to $T=0$ (the solution at $T=0$ being easier than for finite $T$).

\section{Quantum Monte Carlo simulations}
\label{sec_qmc}

\begin{figure}
\includegraphics[width=8cm]{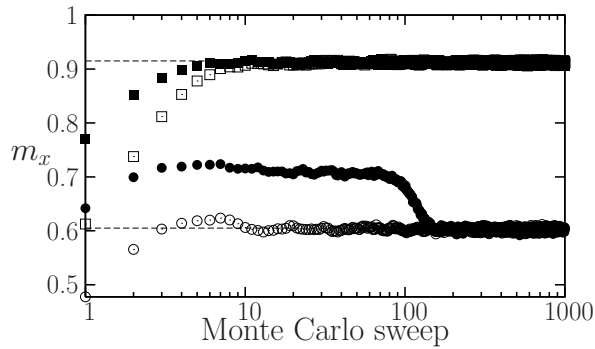}
\caption{Quantum heat bath (full symbols) {\it vs.} Loop Algorithm
(open symbols) for a random regular graph of degree 3 with $10^5$ spins 
at $B=1.6$ (circles), in the ferromagnetic phase, and $B=2.6$ (squares), in the
quantum paramagnetic phase, for $T=1$.  Dashed lines correspond to
the values computed with the cavity method. }
\label{fig_heat}
\end{figure}

\subsection{Methods and algorithms}
The quantum Monte Carlo method  is an important tool that is widely
used to study quantum statistical physics and quantum phase
transitions, especially in the context of lattice models
\cite{ALPS}. In this section, we discuss the application of the
ideas discussed in this paper to quantum Monte Carlo simulations.

\subsubsection{A generic quantum heat bath Monte Carlo scheme}
The procedure we discussed in
sec.~\ref{sec_ct} to generate the ``continuous time'' spin
configurations can be actually used directly as a {\it continuous time
quantum heat-bath algorithm}.
Once the procedure that generates the
new continuous time configuration given the local field trajectory is
available, the implementation of the Monte Carlo simulation is rather
straightforward: one just randomly picks a site, computes the local field
trajectory due to its neighbouring spins, 
and generates a new imaginary-time trajectory of
this spin according to the rules discussed in sec.~\ref{sec_ct}.

The advantage of this method is that one can apply it to {\it any}
discrete spin models on {\it any} kind of lattice.  In the case of
ferromagnetic non frustrated interactions, there exist of course a
number of algorithms that allow to considerably speed up the
simulation and this is the subject of the next section. However, for highly
disordered and frustrated systems such as spin glasses, for instance,
no such {\it generic} cluster algorithm exists even at the classical
level (although interesting progresses are being made, see
\cite{CLUSTER_SG1,CLUSTER_SG2,CLUSTER_SG3,CLUSTER_SG4}) and most
classical simulations \cite{Enzo} rely on the Metropolis or the heat
bath algorithm and the parallel tempering technique \cite{PT_SG}. In
the case of quantum spin glasses, to the best of our knowledge, only
the usual discrete time Suzuki-Trotter decomposition has been tried
\cite{SG_QUANTUM1,SG_QUANTUM2,SG_QUANTUM3}.  Another important case
where no loop algorithm is known is the one of multi-spin
interactions, where the problems are defined on a factor graph 
(see section~\ref{sec_qcav_generic}), which is
very common in the context of constraint satisfaction problems such as
satisfiability or in coding theory.  There is thus a clear need of an
efficient {\it generic} heat-bath strategy when no cluster or loop
algorithm exists, that would result in a substantial gain of
simulation time.

In Fig.~\ref{fig_heat}, we show the first application of our continuous
time quantum heat-bath algorithm to the case of the ferromagnetic model on
a regular random graph of connectivity $3$, as studied in the rest of
the present paper.  The results of our algorithm converge fastly, with
respect to the number of Monte Carlo sweeps, to the asymptotic ones for
very large samples (i.e. $N=10^5$ spins). In the paramagnetic phase,
the convergence is even {\it faster} than for the loop algorithm
discussed in the next section. The application of these ideas and
methods to more complex models is an interesting direction of study.

\subsubsection{Continuous time loop algorithm}

Since we are dealing with a ferromagnetic, non frustrated model, we
expect the usual cluster and loop algorithms
\cite{Beard_Wiese,evertz,LOOPRECENT} to allow for a significant speed
up of the simulation with respect to the heat-bath procedure, even on
random graphs. We have thus implemented the algorithms devised by
Rieger and Kawashima in \cite{qcluster}, which is an adaptation of the
classical Swendsen-Wang \cite{SW} algorithm to quantum systems in continuous
time. The results are compared with the heat-bath method in
Fig.\ref{fig_heat}. Indeed, the loop algorithm performs very well and
better than the heat-bath method close and below the critical point,
and we thus have used this method to compare the results of the cavity
approach with finite size instances.

\subsection{Cavity method versus simulation}

\begin{figure}
\includegraphics[width=8cm]{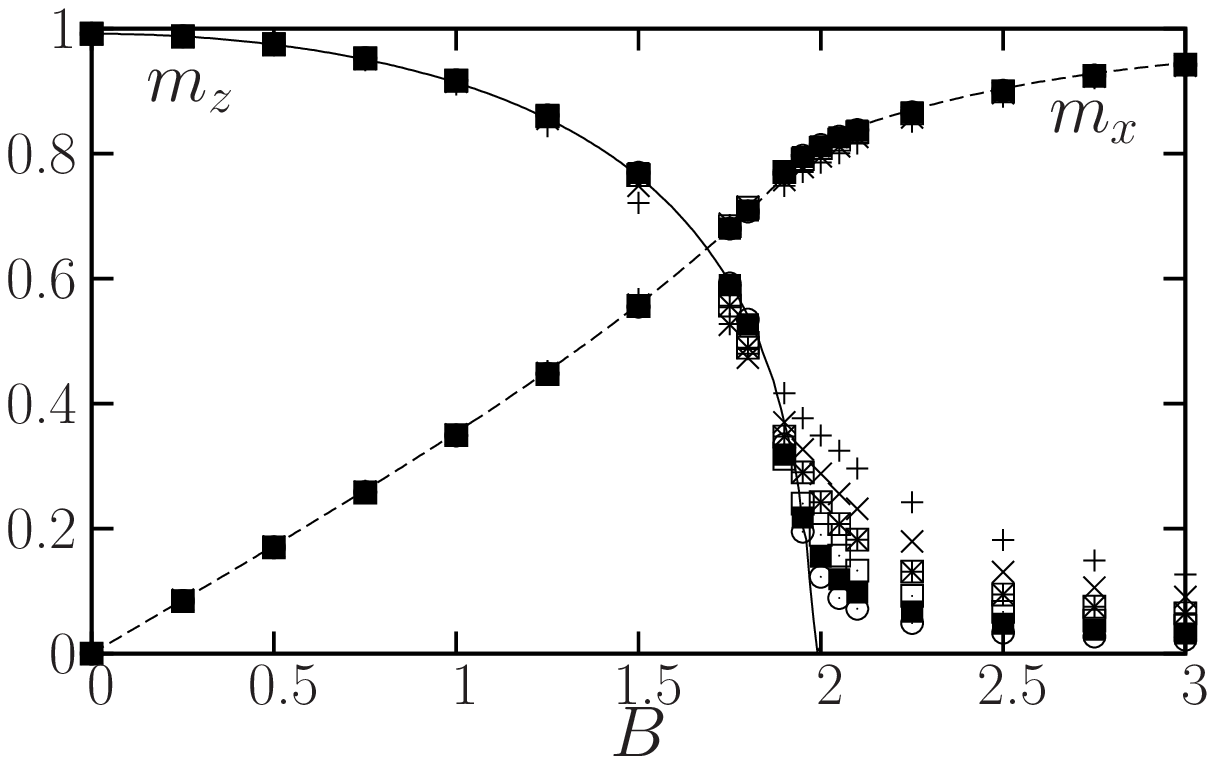}
\includegraphics[width=8cm]{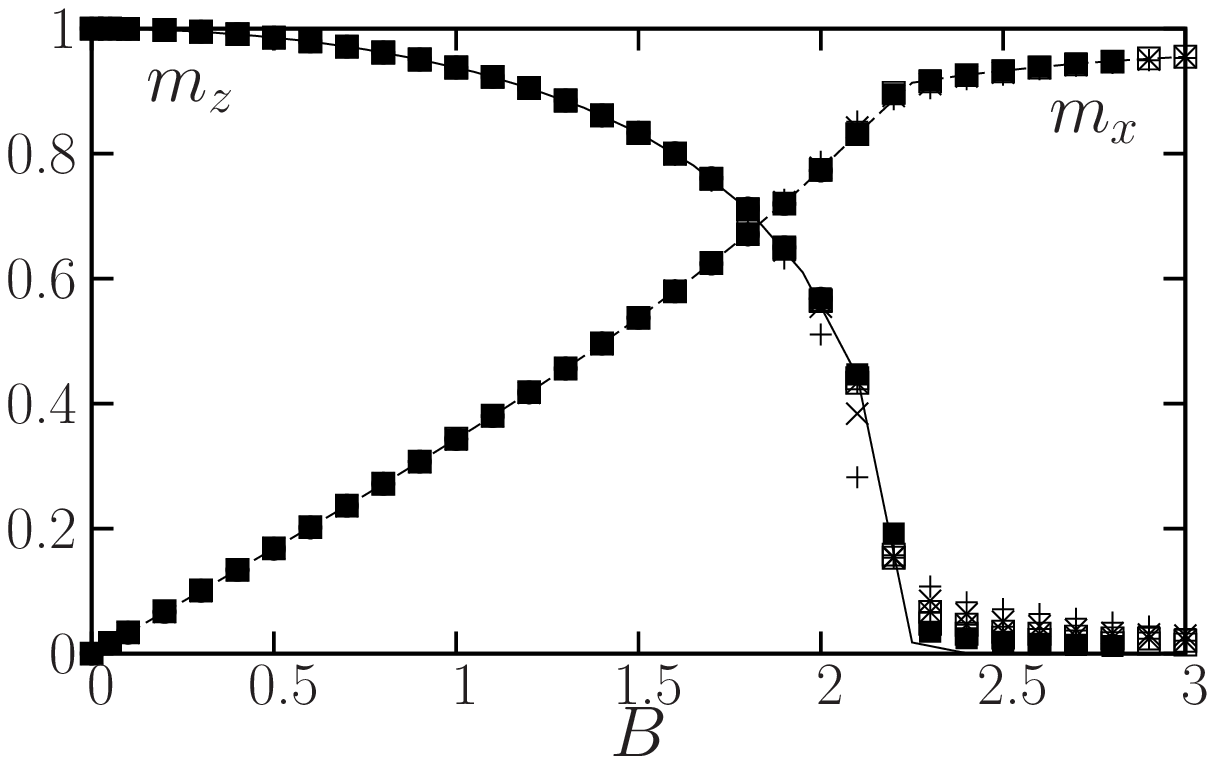}
\caption{Longitudinal and transverse magnetization curves for
temperature $T=1$ (left) and $T=0.1$ (right).  Continuous time cavity
method ($m_z$, solid line and $m_x$, dashed line) 
{\it vs.} Monte Carlo simulations for different
sizes increasing from the top to the bottom
($N=64,128,\ldots,2048$). For the largest size the agreement with the
cavity result is excellent (except close to the critical point).
}
\label{fig_magnetization}
\end{figure}

How does the cavity predictions compare with numerical simulations? To
answer this question, we have performed quantum Monte Carlo
simulations on random regular graphs of $N$ spins and connectivity
$3$.  For large $N$ we expect the results to be self-averaging and
thus we always consider a single instance, and do not perform averages
of many realizations. The results for two different temperature $T=1$
and $T=0.1$ are shown in Fig.\ref{fig_magnetization} where we plot the
longitudinal and transverse magnetization as function of $B$ for
different sizes from $N=64$ to $N=2048$. The agreement with the
asymptotic cavity result is perfect (apart from finite size effects in
$m_z$ for $B \approx B_{\rm c}$, see next paragraph): this demonstrates the
correctness of the approach we have developed in the present paper.

\subsection{Finite size scaling and critical exponent}
\label{sec_FSS}

The transition between the ferromagnetic and the paramagnetic phase is of
second order; below the threshold $B_c(T)$ the longitudinal magnetization
grows with an exponent $\beta$, i.e. $m_z \propto \left( B_{\rm c}-B
\right)^{\beta}$, in the thermodynamic limit. We have used finite size scaling
techniques~\cite{BINDER,FSS} to analyze our data in the neighborhood of the 
transition and to check the mean-field value of the exponent $\beta=1/2$. 
Let us first briefly recall the basic idea of the finite size scaling method
and the way in which it has to be amended for mean-field models. 

\begin{figure}
\includegraphics[width=8.5cm]{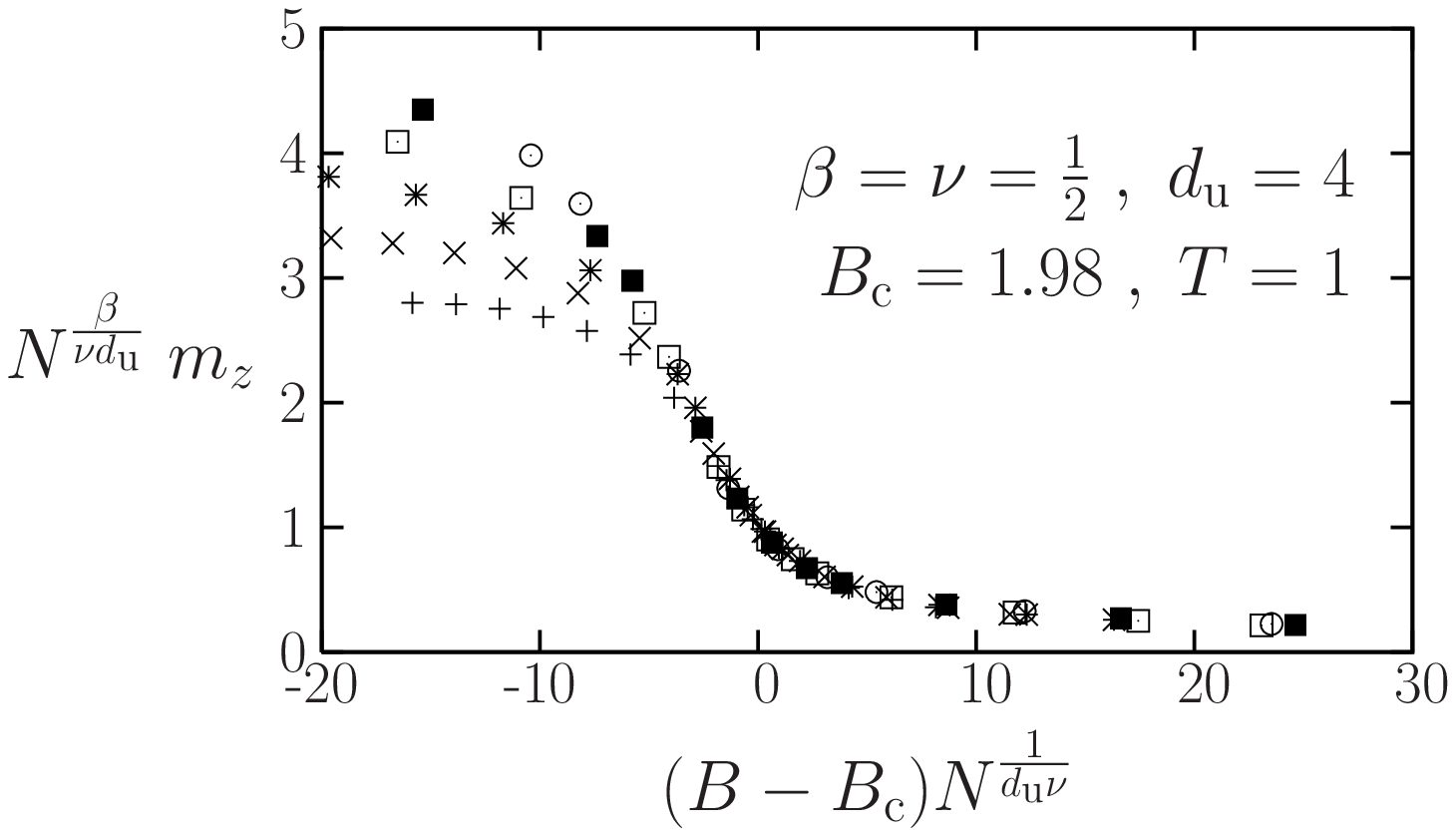}
\includegraphics[width=7.5cm]{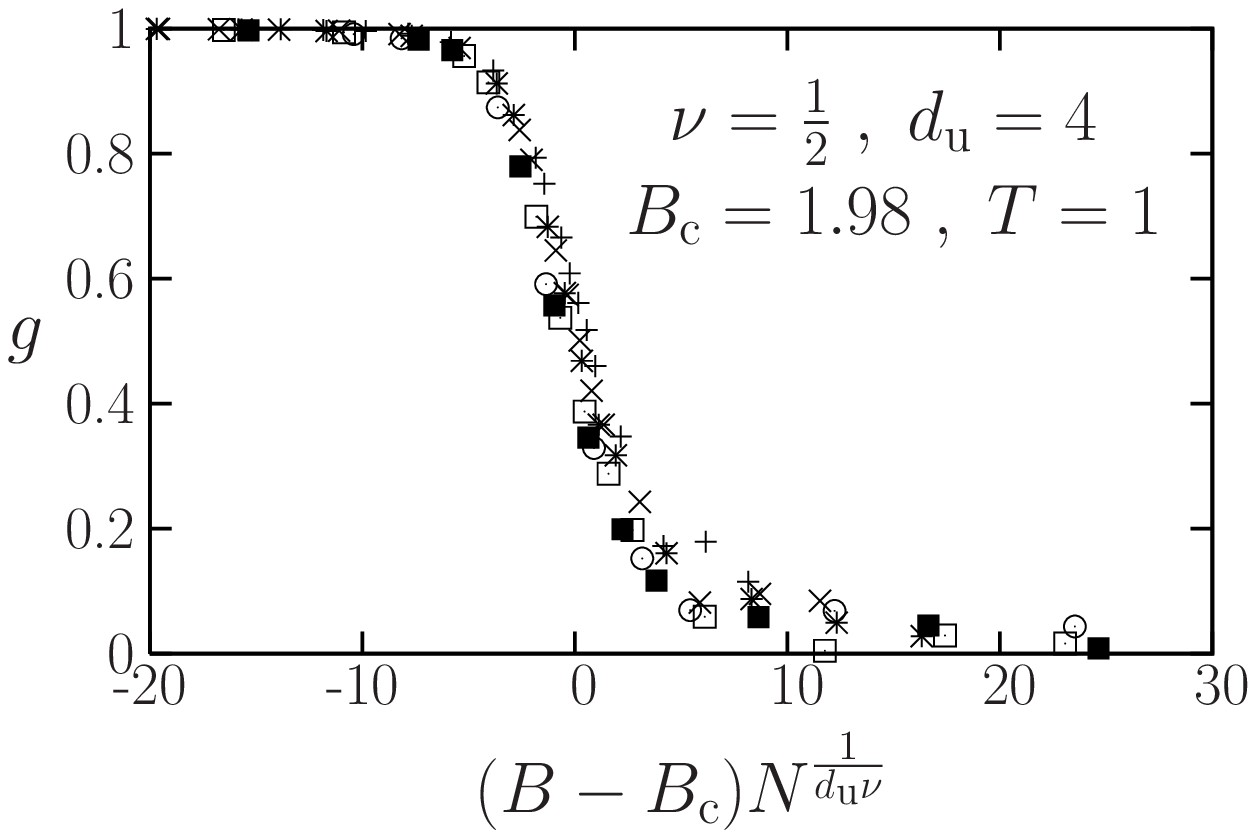}
\caption{Finite size scaling analysis of the Monte-Carlo data. 
Left: rescaling of the 
longitudinal magnetization for random graphs of different sizes 
(increasing from the top to the bottom) $N=64,128,\ldots,2048$.
 Right: Binder parameter for the same sizes.}
\label{fig_binder}
\end{figure}

In a generic infinite size $d$-dimensional model, in the vicinity of a second 
order phase transition driven by a parameter denoted $B$, the correlation 
length diverges as $\xi \propto | B-B_{\rm c} |^{-\nu}$. For a system of finite 
extent $L$ the observables depends on the size through a scaling function of 
the ratio $L/\xi$. This has to be corrected for dimensions $d$ larger than the
upper critical dimension $d_{\rm u}$ of the considered universality class, or
when the model lacks any underlying finite dimensional structure. In that case
the scaling function is 
found~\cite{BREZIN,Botet,YOUNG,EXEMPLE_FFS1,EXEMPLE_FFS2} to depend on the 
size $N$ (total number of 
degrees of freedom, equivalent to $L^d$ of a $d$-dimensional model) through
$N^{1/(d_{\rm u} \nu)}(B-B_{\rm c})$, where $\nu$ takes its mean-field value in 
the universality class under investigation. More explicitly the scaling forms
of the longitudinal magnetization and of the Binder cumulant 
$g=\frac 12 \left(3-\frac{<m_z^4>}{<m_z^2>^2}\right)$
read
\begin{eqnarray} 
  m_z(B, N ) &=& N^{-\beta/d_u \nu} \widetilde m\left(N^{1/(d_u \nu)} \left(B-B_c\right)\right) \ ,\nonumber \\
g(B, N ) &=& \widetilde
g\left(N^{1/(d_u \nu)} \left(B-B_c\right)\right)~.
\end{eqnarray}
We present in Fig.~\ref{fig_binder} the results of such an analysis for
our Monte-Carlo data obtained at temperature $T=1$. At this positive 
temperature the critical behavior of a
quantum $d$-dimensional system is equivalent to a classical
$d$-dimensional Ising model~\cite{suzuki,Sachdev}, which implies $d_{\rm u}=4$
and $\nu=1/2$. Using the cavity method predictions $B_{\rm c}(T=1)=1.98$ and
$\beta=1/2$ we found a very good data collapse (see Fig.~\ref{fig_binder}),
which confirms the validity of these values of $\beta$ and $B_{\rm c}$.

\begin{figure}
\includegraphics[width=8.5cm]{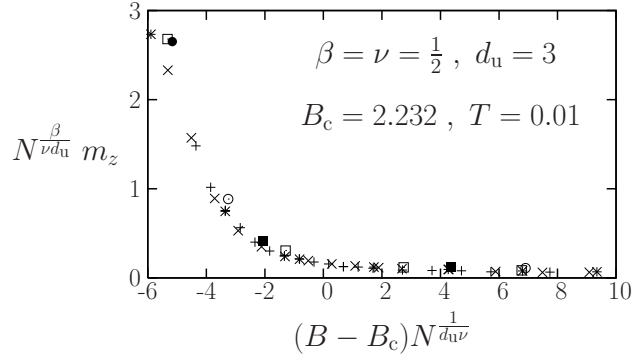}
\caption{Finite size scaling analysis of the longitudinal magnetization
at $T=0.01$, for sizes $N=32,64,\ldots,2048$.}
\label{fig_fss_Tsmall}
\end{figure}

To close this section, let us discuss the value of the critical exponent 
$\beta$ at $T=0$. We believe that because of the mean-field nature of the
Bethe lattice model, $\beta$ keeps the same value $1/2$ at positive and zero
temperatures. This is indeed what happens for the Curie-Weiss model, 
see Appendix~\ref{app_qcw}. A simple argument is the following.
For ferromagnetic models the Suzuki-Trotter formalism
and numerical simulations suggest that the critical behaviour of the
quantum $d$-dimensional model at zero temperature corresponds to the one of 
the classical model in $d+1$ dimensions~\cite{suzuki,Sachdev,qcluster}. 
Hence if a model behaves in a mean-field way at positive temperature it should
also do so at zero temperature, having formally gained one more spatial 
dimension. To give further credit to our theses we performed 
quantum Monte-Carlo
simulations at very small temperatures, $T=0.01$. The plot of 
Fig.~\ref{fig_fss_Tsmall} shows again a very good collapse with the value 
$\beta=1/2$, using the zero temperature critical value of the transverse
field extracted from the cavity computation, $B_{\rm c}=2.232$. This collapse
has been obtained with $d_{\rm u}=3$; indeed at such a low temperature we are
well inside the low temperature regime where the transition is truly quantum,
hence the shift of the upper critical dimension to account for the imaginary 
time supplementary dimension.

A different value of the zero temperature exponent, $\beta = 0.41$, 
has been obtained in~\cite{qbl} using the matrix product state ansatz
for the description of the groundstate. Though we cannot rule out
the possibility that a crossover occurs at temperatures even lower than 
$T=0.01$, we believe that $\beta=1/2$ down to $T=0$, in accordance with 
the mean-field nature of the Bethe lattice, and that the different
value reported in~\cite{qbl} might be due to the truncation of the matrix 
product state ansatz.

\section{The generic replica symmetric quantum cavity method}
\label{sec_qcav_generic}

In this final Section we give a more generic description of the
replica symmetric quantum cavity method. As the main ideas should have already
been conveyed by the example of the ferromagnet on the random regular graphs
we shall mainly emphasize the differences and complications that arise in more
general models.

\begin{figure}
\includegraphics[width=6cm]{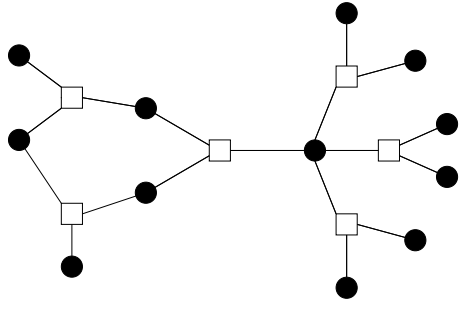}
\caption{An example of a factor graph}
\label{fig_fgraph}
\end{figure}

Let us consider the class of Ising spin models whose classical energy $E(\us)$ 
can be decomposed into the sum of $M$ interaction terms $a=1,\dots,M$,
each of them depending on a finite number of spins,
\begin{equation}
E(\us) = \sum_{a=1}^M \ve(\us_{\partial a} ,J_a) \ .
\end{equation}
In this expression $\partial a$ is the subset of spin indices the $a$-th 
interaction effectively depends on, 
$\us_{\partial a}=(\s_i : i \in \partial a)$ is a shorthand for the 
configuration of those spins, and $J_a$ denotes coupling constants that might
appear in the definition of the $a$-th interaction term. This decomposition
is conveniently represented as a factor graph~\cite{fgraph}, i.e. a graph with 
two kind of vertices (see Fig.~\ref{fig_fgraph} for an illustration): 
squares stand for the interactions $a=1,\dots,M$, while circles represent the
variables $i=1,\dots,N$. An edge is drawn between an interaction $a$ and a
variable $i$ whenever $a$ depends on $i$, in other words whenever 
$i\in \partial a$. In this graphical representation $\partial a$ is thus the 
set of nodes adjacent to $a$. Similarly we shall denote $\partial i$ the set
of neighbors of $i$, that is all the interactions that depend on $\s_i$.
The extended Ising model defined in Eq.~(\ref{eq_def_mu_ubs}) preserves the
topology of the interactions of the classical energy $E(\us)$, the latter
being reproduced identically in the various imaginary time slices,
\begin{eqnarray}
\tE(\ubs) &=& \sum_{a=1}^M \tve(\ubs_{\partial a},J_a) \ , \nonumber \\
\tve(\ubs_{\partial a},J_a) &=& \frac{1}{\Ns} \sum_{\alpha=1}^{\Ns}
\ve(\us_{\partial a}^\alpha,J_a) \ ,
\end{eqnarray}
while the transverse weights $w(\bs_i)$ are local in the spin indices $i$.

As we did on the ferromagnet example we first state the solution of this
extended model on a tree. The difference is that we have now two kind of
``messages'', from variables to interactions and vice versa.
Let us denote $\mu_{a \to i}(\bs_i)$ the
probability law of $\bs_i$ in the model corresponding to the factor graph
where all interactions but $a$ have been removed in the neighborhood of $i$,
and similarly $\mu_{i \to a}(\bs_i)$ for the factor graph with only $a$ 
removed from $\partial i$. These quantities obey the following recursion
equations,
\begin{eqnarray}
\mu_{i \to a}(\bs_i) &=& \frac{1}{z_{i \to a}} w(\bs_i) 
\prod_{b \in \partial i \setminus a} \mu_{b \to i}(\bs_i)  \ , \label{eq_bp1}\\
\mu_{a \to i}(\bs_i) &=& \frac{1}{z_{a \to i}} 
\sum_{\{\bs_j\}_{j \in \partial a \setminus i} } 
e^{-\beta \tve(\ubs_{\partial a},J_a )} \prod_{j \in \partial a \setminus i}
\mu_{j \to a}(\bs_j) \ ,
\nonumber
\end{eqnarray}
on all edges of the factor graph, the various constants $z$ being normalization
factors. The reader will easily verify that these equations reduce to 
Eq.~(\ref{eq_ferro_mu}) in the ferromagnetic case with 
$\ve(\s_i,\s_j)=- \s_i \s_j$; in this case we could eliminate one type 
of message (from interactions to variables), as the interactions were 
only pairwise.

\begin{figure}
\includegraphics[width=6cm]{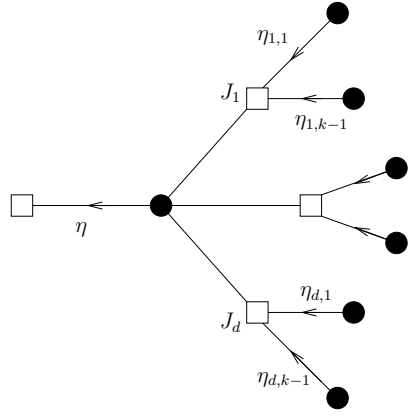}
\caption{A pictorial representation of Eq.~(\ref{eq_recurs}).}
\label{fig_eq_msg}
\end{figure}

We shall now consider ensembles of random factor graphs, denoting $\E[\cdot]$
the expectation over the distribution of the graphs and coupling constants.
We assume all interaction nodes to involve a fixed number $k$ of variables
(the ferromagnet corresponded to $k=2$),
while the degree distribution of the variables is specified by a probability
law $q_d$ over the positive integers (all random graphs verifying these 
constraints are equiprobable in the ensemble; the lack of a finite dimensional
a priori structure is the origin of the mean-field character of this family of
models). We shall denote 
$\gamma k = \sum_d d q_d$ the average variable degree. Moreover the 
$M = \gamma N$ coupling constants $J_a$ are drawn in an identical, 
independent way for each of the interactions.
Suppose the recursion
equations (\ref{eq_bp1}) are solved on a factor graph sampled at
random from the ensemble under consideration, and that an edge from a variable
to an interaction, call it $i \to a$, is chosen uniformly at random. The
probability law $\mu_{i \to a}$ it bears is itself a random variable, denoted
$\eta$ in the following, due to
the random choices of the graph and of its coupling constants. This random
variable can be related to its neighboring equivalents by the local relations
(\ref{eq_bp1}). Suppose that $i$ has $d$ adjacent interactions
apart from $a$ (see Fig.~\ref{fig_eq_msg} for an illustration). Then one has,
from (\ref{eq_bp1}),
\begin{widetext}
\begin{equation}
\eta(\bs) = \frac{1}{z} w(\bs) 
\sum_{\{ \bs_{a,i} \}_{a\in[1,d]}^{i\in[1,k-1]}} 
\left( \prod_{a,i} \eta_{a,i}(\bs_{a,i}) \right) \ 
e^{-\beta \sum_{a=1}^d \tve(\bs,\bs_{a,1},\dots,\bs_{a,k-1},J_a ) } \ ,
\label{eq_recurs}
\end{equation}
\end{widetext}
where the $\eta_{a,i}$ are the $d(k-1)$ probability laws that determine 
$\eta$. The hypothesis of the replica symmetric cavity method is to assume
that the above form is correct in spite of the graph not being globally a tree,
and that the $\eta_{a,i}$ are independent identically distributed copies of the
random variable $\eta$, which is written in formula as
\begin{equation}
\eta \eqd 
f(\eta_{1,1},\dots,\eta_{1,k-1},\dots,\eta_{d,1},\dots,\eta_{d,k-1},
J_1,\dots,J_d) \ , 
\label{eq_distrib}
\end{equation}
the symbol $\eqd$ denoting identity in distribution of random variables, 
and the function $f$ being an abbreviation of the right-hand-side of 
(\ref{eq_recurs}).
Note that the connectivity random variable $d$ is not distributed according
to $q_d$ but rather $\tq_d = (d+1) q_{d+1} / \gamma k $: as one uniformly 
choose a random edge of the graph, and not a random site, this sampling
favours larger connectivity variables. For the ferromagnet on the random
regular graph one had $q_d=\delta_{d,l+1}$ while $\tq_d=\delta_{d,l}$.

The physical observables of the system can be computed from the solution
of this distributional equation. The thermodynamic limit of the free-energy 
per site is found to be 
\begin{widetext}
\begin{eqnarray}
-\beta f = \lim_{N \to \infty} \frac{1}{N} \E[\ln Z] 
&=& \E\left[\ln\left(   
\sum_{\bs, \{ \bs_{a,i} \}_{a\in[1,d]}^{i\in[1,k-1]}} 
\left( \prod_{a,i} \eta_{a,i}(\bs_{a,i}) \right) w(\bs)
e^{-\beta \sum_{a=1}^d \tve(\bs,\bs_{a,1},\dots,\bs_{a,k-1},J_a ) }  \right)  
\right] \nonumber \\
&-& \gamma (k-1) \E\left[\ln \left( \sum_{\bs_1,\dots,\bs_k}  
\left( \prod_i \eta_i(\bs_i) \right)
e^{-\beta \tve(\bs_1,\dots,\bs_k,J ) }
\right) \right]  \ , \label{eq_f}
\end{eqnarray}
\end{widetext}
where the expectations of the left-hand-side are over the choice of the random
factor graphs, and for the right-hand-side over independent copies of the 
random variable $\eta$ and the coupling constants $J$, while in the first
term $d$ is drawn from the law $q_d$. This is a generalization of
the expression (\ref{eq_qbl_f_var}) found for the ferromagnet. Similarly
the average marginal of the law 
(\ref{eq_def_mu_ubs}) for an arbitrary site reads
\begin{widetext}
\begin{equation}
\mu(\bs) = \E\left[ \frac{1}{z} w(\bs) 
\sum_{\{ \bs_{a,i} \}_{a\in[1,d]}^{i\in[1,k-1]}} 
\left( \prod_{a,i} \eta_{a,i}(\bs_{a,i}) \right) \ 
e^{-\beta \sum_{a=1}^d \tve(\bs,\bs_{a,1},\dots,\bs_{a,k-1},J_a ) }  \right] 
\label{eq_marginal}
\end{equation}
\end{widetext}
with $d$ drawn from $q_d$, while the average marginal law of the $k$ 
variables in an arbitrary interaction is
\begin{equation}
\mu(\bs_1,\dots,\bs_k) = \E \left[\frac{1}{z} 
\left( \prod_i \eta_i(\bs_i) \right)
e^{-\beta \tve(\bs_1,\dots,\bs_k,J ) }  \right] \ ,
\end{equation}
the factors $1/z$ in these last two equations ensuring their normalization
(cf. Eq.~(\ref{eq_marginal_qbl}) for the regular ferromagnet).

Let us now discuss a method of resolution of the distributional equation
(\ref{eq_distrib}), consisting in encoding the distribution of
$\eta$ by a sample (or population) of a large number $\cal N$ of 
representatives $\eta_i$. From an arbitrarily initialized population, one
applies iteration steps according to Eq.~(\ref{eq_distrib}): one draws an
integer $d$ from $\tq_d$, $d$ coupling constants $J_1,\dots,J_d$ and
$d(k-1)$ indices in $[1,{\cal N}]$. The right-hand-side of 
Eq.~(\ref{eq_distrib}) is then computed from the corresponding $d(k-1)$
representants of $\eta$ randomly chosen in the population, and the resulting
$\eta$ is used to replace one discarded element of the population. The 
iteration of these steps brings the population close to a fixed point of
the distributional equation; then physical observables like the average
free-energy (see Eq.~(\ref{eq_f})) can be computed, evaluating the 
expectations over the random variable $\eta$ as a sampling from the 
approximate representation provided by the finite population. Compared
with the classical replica symmetric cavity method the difficulty is
that each of the $\eta_i$ in this population is itself a probability 
distribution over configurations of spin rings, or in the continuous limit
over spin trajectories. It is however possible to apply the trick explained
in the simpler case of the ferromagnet, and encode each of the $\eta_i$ as
a sample of $\Nt$ spin trajectories. Let us introduce some further notations
in order to give an explicit form of the updating procedure of the population
of $\eta_i$'s.

The dependence of $\ve(\cdot,J)$ on one of
the Ising spins can always be parameterized through two functions $u$ and
$v$ as
\begin{eqnarray}
\ve(\s,\s_1,\dots,\s_{k-1},J) = &-& \s \ u(\s_1,\dots,\s_{k-1},J)
+ 
v(\s_1,\dots,\s_{k-1},J) \ ,
\end{eqnarray}
$u$ playing the role of an effective magnetic field acting on $\s$.
For $\Ns$-fold replicated variables we define
\begin{widetext}
\begin{eqnarray}
\bu(\bs_1,\dots,\bs_{k-1},J) &=& 
(u(\s_1^1,\dots,\s_{k-1}^1,J),\dots,u(\s_1^\Ns,\dots,\s_{k-1}^\Ns,J) \ , 
\label{eq_def_bu}
\\
\tv(\bs_1,\dots,\bs_{k-1},J) &=& 
\frac{1}{\Ns} \sum_{\alpha=1}^{\Ns} v(\s_1^\alpha,\dots,\s_{k-1}^\alpha,J)
\label{eq_def_bv}
\end{eqnarray}
\end{widetext}
Using the definitions in Eq.~(\ref{eq_def_p}), we can then rewrite Eq.~(\ref{eq_recurs}) as
\begin{widetext}
\begin{eqnarray}
\eta(\bs) &=& \sum_{\{ \bs_{a,i} \}_{a\in[1,d]}^{i\in[1,k-1]}} 
\left( \prod_{a,i} \eta_{a,i}(\bs_{a,i}) \right)
\ p(\bs |\bh(\{\bs_{a,i} , J_a \})  ) 
\frac{z(\{ \bs_{i,a}, J_a\})}{z} \qquad \text{with}  \nonumber \\
\bh(\{\bs_{a,i} , J_a \}) &=& \sum_{a=1}^d \bu(\bs_{a,1},\dots,\bs_{a,k-1},J_a)
\ , \qquad
z(\{ \bs_{a,i}, J_a \}) = \Z(\bh(\{\bs_{a,i} , J_a \})) 
e^{-\beta \sum_{a=1}^d \tv(\bs_{a,1},\dots,\bs_{a,k-1},J_a)} \ .
\end{eqnarray}
\end{widetext}
We can thus apply the sampling procedure explained in Sec.~\ref{sec_qbl_rep},
the only change being the different definition of the effective
field trajectory $\bh$, and the inclusion of a contribution arising from
the function $v$ in the weights of the generated spin trajectories.
Let us emphasize that at variance with the classical cavity method, here we 
have to deal with a population of population of spin trajectories (of total
size ${\cal N} \times \Nt$) already at the replica symmetric level.

\section{Conclusions}
\label{sec_conclusions}

The explicit procedure for the construction of continuous time spin
trajectories presented in Sec.~\ref{sec_ct} allowed us to make progresses both
on the analytical side, with an improvement over the discretized time cavity
method of~\cite{qc_first} and the static approximation of~\cite{qksat}, 
and on the numerical simulations side, with a
generic quantum Monte Carlo procedure for spin-1/2 models in transverse
field. The ferromagnetic model we studied in this article is only the
simplest of a large family that can be tackled, at the price of an heavier
computational cost, with the same methods. We believe that its study was in 
any case worthwhile from a methodological point of view: its simplicity
permitted a complete resolution of the quantum cavity equations, which
(i) showed a perfect agreement with Monte Carlo simulations, hence giving 
credit to the conjecture that the replica symmetric cavity method leads
to exact results in the thermodynamic limit for sparse mean-field ferromagnetic systems, as
was proved for classical models in~\cite{ferro_rigorous}; (ii) allowed to
test quantitatively some approximate treatments (finite $\Ns$ and static
approximation).

The more interesting cases we plan
to address in the future by mean of the cavity method
will involve fluctuating connectivities, in order to
study the role of these local fluctuations on the critical behaviour of the
models, glassy phases at low temperatures (a particularly motivating case will
be the regular multispin ferromagnet studied at the classical level
in~\cite{ferroglass}), and models related to quantum computing issues, as the
random $k$-satisfiability 
model in a transverse field~\cite{qksat}. Note that already the
replica symmetric treatment of these models will involve a population of
populations of spin trajectories, as explained in Sec.~\ref{sec_qcav_generic},
which will make an exact treatment of the one step replica symmetry breaking
version of the cavity method extremely challenging. We hope however that the
better control of the approximative treatments we gained on the simple 
ferromagnet will help to devise appropriate approximate strategies to handle 
these cases.

At the same time, we proposed a heat bath Monte Carlo
strategy that should be applicable to general spin-1/2 models. The
performance of this algorithm should be tested on frustrated models
for which cluster algorithms are not easy to implement, and we
expect that in such cases the heat bath method could give interesting
results.

The strategy developed above become inefficient at very low
temperatures, because the spins jump many times along a trajectory
and the amount of information needed to encode $\eta(\bs)$ by
a population of trajectories becomes very large. Therefore
an important directions of research would be to look for possible 
simplifications of the formalism in the limit $\beta \to \infty$.
Moreover, one could use the recursion equations (\ref{eq_bp1}) 
on a given sample, thus
constructing a quantum belief propagation algorithm~\cite{qbp}.
Finally, it is worth to note that the procedure described in
section~\ref{sec_ct}, on which our results are based,
is in principle not
restricted to spin-1/2 models. It should be possible to generalize
it to any model built on discrete degrees of 
freedom~\cite{Farhi_Gutmann}, for instance
hard-core bosons or higher spin models.

\acknowledgments

We wish to thank G.~Biroli, W.~Krauth, M.~M\'ezard, M.~M\"uller, H.~Rieger,
A.~Scardicchio, S.~Sondhi, P.~Young, M.~Tarzia, L.~Zdeborov\'a for discussions and
comments on a first version of the paper.

\appendix

\section{Quantum Curie-Weiss model}
\label{app_qcw}

\begin{figure}
\includegraphics[width=8cm]{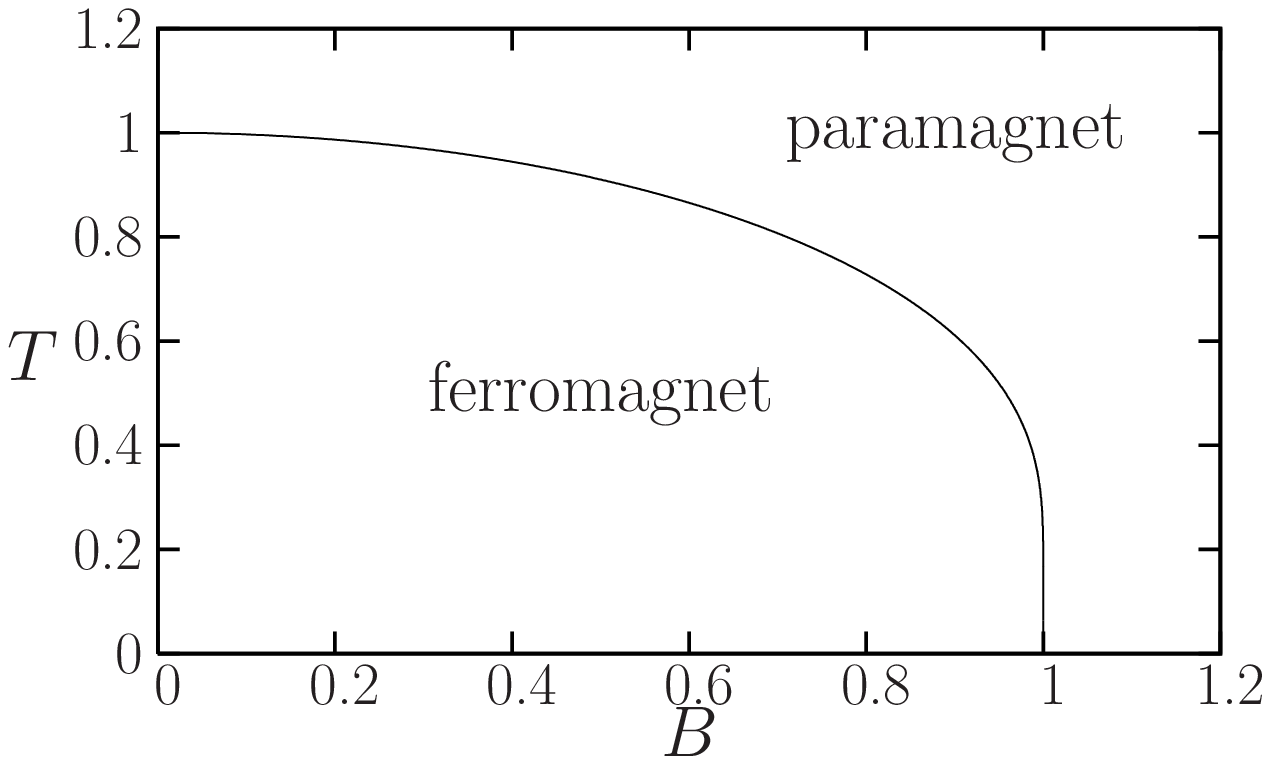}
\includegraphics[width=8cm]{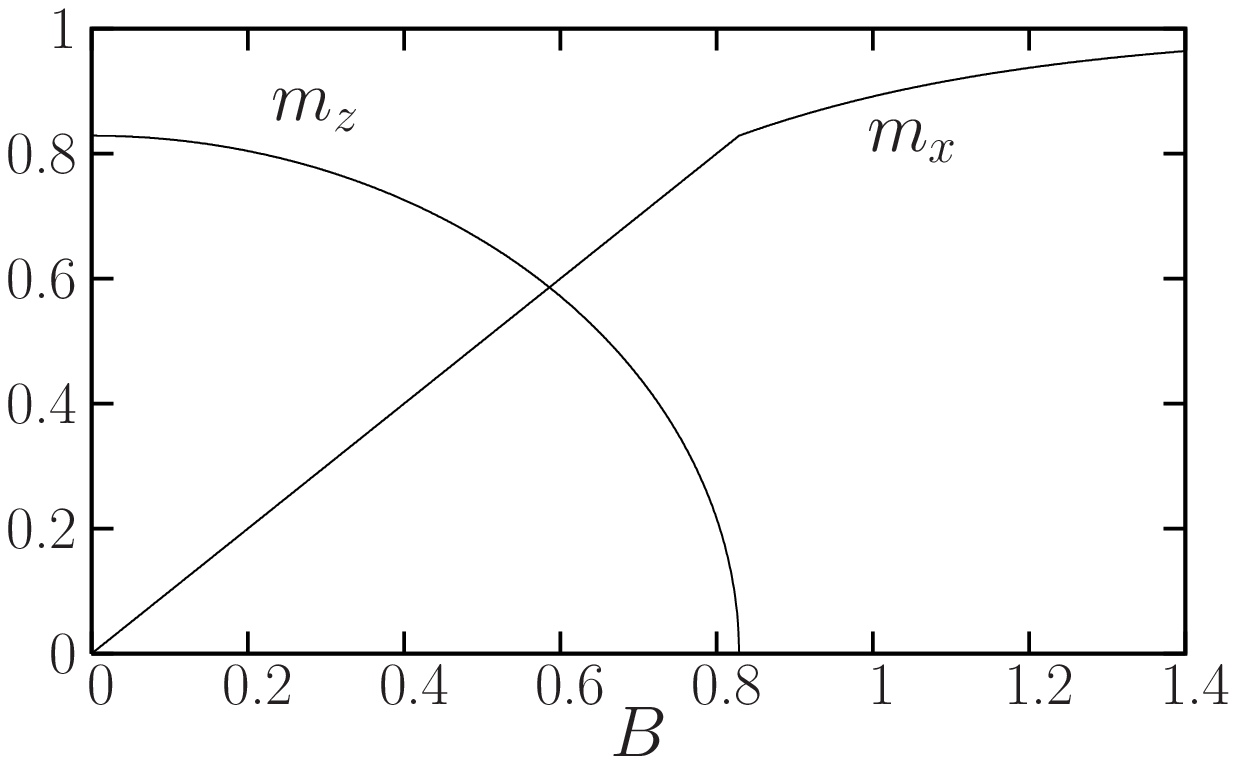}
\caption{Left: the phase diagram of the quantum Curie-Weiss model. Right:
longitudinal and transverse magnetizations as a function of the transverse 
field, for $T=0.7$.}
\label{fig_cw}
\end{figure}

This Appendix is devoted to a study of the simplest quantum mean-field
ferromagnet, namely the fully-connected Curie-Weiss quantum
model~\cite{qcw_1,qcw_2,qcw_math}. The model is defined by the Hamiltonian
\begin{equation}
\hH = - \frac{J}{2N} \sum_{i,j=1}^N \s_i^z \s_j^z - B \sum_{i=1}^N \s_i^x 
- h \sum_{i=1}^N \s_i^z \ ,
\end{equation}
where the scaling of the coupling constant is chosen appropriately to make
the thermodynamic limit well-defined. At variance with the Bethe lattice model
here each spin interacts with all others. Applying the Suzuki-Trotter 
decomposition described in Sec.~\ref{sec_st} leads for a finite number 
$\Ns$ of Suzuki-Trotter slices to the following expression of the partition
function:
\begin{widetext}
\begin{equation}
Z= \sum_{\ubs} 
\left(\prod_{i=1}^N w(\bs_i) \exp\left[ \frac{\beta h}{\Ns} 
\sum_{\alpha=1}^\Ns \s_i^\alpha \right]
\right) \exp \left[ \frac{\beta J}{2 N \Ns}
\sum_{\alpha=1}^\Ns \sum_{i,j=1}^N \s_i^\alpha \s_j^\alpha \right] \ .
\end{equation}
\end{widetext}
We then perform $\Ns$ Hubbard-Stratanovitch transformations to disentangle
the quadratic terms and obtain
\begin{widetext}
\begin{equation}
Z = \left(\frac{\beta J N}{2\pi \Ns} \right)^{\frac{\Ns}{2}}
\int \prod_{\alpha=1}^\Ns \de m^\alpha
\exp\left[
-N \frac{\beta J}{2} \frac{1}{\Ns} \sum_{\alpha=1}^\Ns (m^\alpha)^2 + 
N \ln \text{Tr} \left( \prod_\alpha e^{\frac{\beta}{\Ns}(h+J m^\alpha) \s^z } 
e^{\frac{\beta}{\Ns} B \s^x } \right) \right] \ .
\end{equation}
\end{widetext}
Evaluating these integrals by the saddle-point method in the thermodynamic 
limit and selecting the cyclically invariant saddle point yields
\begin{equation}
\lim_{N \to \infty} \frac{1}{N} \ln Z = \sup_{m}
\left[
-\frac{\beta J}{2} m^2 
+ \ln \text{Tr} \left( \left( e^{\frac{\beta}{\Ns}(h+J m) \s^z } 
e^{\frac{\beta}{\Ns} B \s^x } \right)^\Ns \right) \right] \ .
\label{eq_cw_before_f}
\end{equation}
This can be further simplified if the $\Ns \to \infty$ limit is performed
afterwards and yields for the free-energy per site
\begin{equation}
f = \inf_{m} \left[
\frac{J}{2} m^2 
 -\frac{1}{\beta} \ln \left(2 \cosh \left(\beta \sqrt{(h+J m)^2 +B^2 } \right)
\right)\right] \ .
\label{eq_cw_f}
\end{equation}
Using the variational character of this expression
the longitudinal and transverse magnetizations (i.e. $m_z$
and $m_x$) can be obtained by taking the explicit
derivatives with respect to $h$ and $B$ respectively, 
\begin{widetext}
\begin{eqnarray}
m_z \equiv \langle \s_i^z\rangle &=& \frac{h+J m}{\sqrt{(h+J m)^2 +B^2 }}
\tanh \left(\beta \sqrt{(h+J m)^2 +B^2 } \right) \ , \label{eq_cw_mz} \\
m_x \equiv \langle \s_i^x\rangle &=& \frac{B}{\sqrt{(h+J m)^2 +B^2 }}
\tanh \left(\beta \sqrt{(h+J m)^2 +B^2 } \right) \ ,\label{eq_cw_mx}
\end{eqnarray}
\end{widetext}
where $m$ is taken as the solution of the saddle-point equation. The latter
is easily found to imply that $m = m_z$ at the saddle-point. 
In the following we take $J=1$ to simplify the notations.

In absence of the transverse field ($B=0$) one recovers the classical 
Curie-Weiss model, with $m = m_{\rm cl}(\beta,h)$ solution of 
the traditional equation $m=\tanh(\beta (m+h))$. The ferromagnetic transition
is signaled by the appearance of a non-trivial solution at $h=0$, which is
possible for small enough temperatures, i.e. for 
$\beta > \beta_{\rm c}(B=0)=1$.

Consider now the solutions of the saddle-point equation with $B>0$ and $h=0$.
The paramagnetic solution $m=0$ exists for all temperatures and transverse 
fields. For a strictly positive solution $m$ the saddle-point equation 
reduces to $\sqrt{m^2 + B^2} =\tanh(\beta \sqrt{m^2 + B^2} )$, that is
$m(\beta,B,h=0) = \sqrt{m_{\rm cl}(\beta,h=0)^2 - B^2}$. This is possible
only for small enough temperatures and transverse fields, such that the
argument of the square root remains positive. The line of transition in the
$(B,T)$ plane is such that $B_{\rm c}(\beta)=m_{\rm cl}(\beta,h=0)$, see
the top panel of Fig.~\ref{fig_cw}. Note the vertical slope of the transition
line in the neighborhood of the quantum critical point in $(B=1,T=0)$; in fact
one can perform an asymptotic expansion in this region, to show that
\begin{equation}
B_{\rm c}(T) \underset{T \to 0}{\sim} 1 - 2 e^{-2 \beta} \ ,
\label{eq_BcT0_qcw}
\end{equation}
or equivalently,
\begin{equation}
T_{\rm c}(B) \underset{B \to 1}{\sim} 
\frac{1}{\ln \left(\frac{1}{\sqrt{1-B}} \right)} 
\ .
\end{equation}
The bottom panel of Fig.~\ref{fig_cw} shows the evolution, as a function of 
the transverse field, 
of the longitudinal and transverse magnetizations at a temperature smaller
than the classical critical temperature $T_{\rm c}(B=0)=1$. The transverse
magnetization is continuous but its derivative has a finite jump at the 
transition. One can indeed show from Eq.~(\ref{eq_cw_mx}) that
$m_x = B$ in the ferromagnetic phase, while
$m_x=\tanh(\beta B)$ for the paramagnetic one.

Another thermodynamic quantity easily computed for the Curie-Weiss model is
the ground-state energy per spin, $e_{\rm gs}(B)$, obtained from 
(\ref{eq_cw_f}) in the limit $\beta \to \infty$. One finds 
\begin{equation}
e_{\rm gs}(B) = \begin{cases} -\frac{1}{2} (1+B^2) 
& \text{for} \ B \le B_{\rm c} = 1 \\
- B  & \text{for} \ B \ge B_{\rm c} = 1
\end{cases} \ ,
\end{equation}
which shows that $e_{\rm gs}(B)$ and its first derivative are continuous at the
transition, while its second derivative has a finite jump.

Let us finally argue about the scaling of the finite $\Ns$ corrections,
reconsidering the step we took between Eqs.~(\ref{eq_cw_before_f}) and 
(\ref{eq_cw_f}). At the next-to-leading order one would have obtained
\begin{widetext}
\begin{equation}
\Tr \left(\left( e^{\frac{\beta}{\Ns}(h+J m) \s^z } 
e^{\frac{\beta}{\Ns} B \s^x } \right)^\Ns \right) = 
\Tr \left( \exp \left[ \beta(h+Jm)\s^z + \beta B \s^x 
+ \frac{1}{2\Ns} \beta^2 B (h+J m) [\s^z,\s^x]  \right]
 \left(1 + O(\Ns^{-2}) \right) \right) \ .
\nonumber
\end{equation}
\end{widetext}
One can then check explicitely that the eigenvalues of the matrix in
the exponential are not modified at the order $\Ns^{-1}$, hence the 
finite $\Ns$ correction on the observables should be of order $\Ns^{-2}$,
as we observed in the study of the Bethe lattice ferromagnet.

\section{An identity in the continuous time limit}
\label{app_details_ct}
In this appendix we briefly justify the claim made in Sec.~\ref{sec_qbl_rep}
of the equivalence in the continuous time limit of the definition of $\Z(\bh)$
given in (\ref{eq_def_p}) with the one of (\ref{eq_def_Z_bh}).
Before taking the $\Ns \to \infty$ limit, the former reads
\begin{eqnarray}
\Z(\bh) &=& \sum_{\bs} w(\bs) \exp[\beta \ \bs \cdot \bh ] 
= \sum_{\s^1,\dots,\s^\Ns} \prod_{\alpha=1}^\Ns 
\langle \s^\alpha | e^{\frac{\beta}{\Ns} h^\alpha \s^z } e^{\frac{\beta}{\Ns}
  B \s^x} | \s^{\alpha+1} \rangle \ . \nonumber
\end{eqnarray}
We shall denote $\Ns^{(i)} = \Ns \lambda^{(i)}/ \beta $ the lengths of the
constant longitudinal field intervals (see Fig.~\ref{fig_traj_h}), expressed
in number of Suzuki-Trotter slices. Then
\begin{widetext}
\begin{eqnarray}\nonumber
\Z(\bh) &=& \sum_{\s(t^{(0)}),\dots,\s(t^{(p)})} \prod_{i=1}^p 
\langle \s(t^{(i)}) |  
\left(e^{\frac{\beta}{\Ns} h^{(i)} \s^z } 
e^{\frac{\beta}{\Ns} B \s^x}\right)^{\Ns^{(i)}} | \s(t^{(i+1)}) \rangle
\qquad \s(t^{(p+1)})=\s(t^{(0)}) \\
&=& \sum_{\s(t^{(0)}),\dots,\s(t^{(p)})} \prod_{i=0}^p \langle \s(t^{(i)}) |
e^{\lambda^{(i)}(h^{(i)} \s^z + B \s^x ) }
| \s(t^{(i+1)}) \rangle \ .
\end{eqnarray}
\end{widetext}
In the last step we have taken the $\Ns \to \infty$ limit to obtain
the expression of Eq.~(\ref{eq_def_Z_bh}).

\section{The computation of $w_{\rm s}(m)$}
\label{app_ws}
In this appendix we derive the expression (\ref{eq_ws}) for the average
transverse weight in the static approximation. Rewriting the sum over $\bs$
of (\ref{eq_u0}) as a sum over paths in the continuous time limit, one obtains
\begin{widetext}
\begin{eqnarray}
w_{\rm s}(m) &=& \sum_\s \sum_{n=0}^\infty B^{2n} \int_0^\beta \de t_1 
\int_{t_1}^\beta \de t_2 \dots \int_{t_{2n-1}}^\beta \de t_{2 n}
\ \delta\left(m - \s \frac{2t_1 - 2t_2 + \dots - 2t_n + \beta}{\beta} \right) 
\ , \\
&=& \delta(m-1) + \delta(m+1) + \sum_\s \sum_{n=1}^\infty (B\beta)^{2n} 
\int_0^1 \de x_1 
\int_{x_1}^1 \de x_2 \dots \int_{x_{2n-1}}^1 \de x_{2 n}
\ \delta\left(2x_1 - 2x_2 + \dots - 2x_n  +1 - m \s \right) \ ,
\nonumber
\end{eqnarray}
\end{widetext}
where in the second line we have isolated the contribution of the constant
trajectories, and changed variables from $t_i$ to $x_i =t_i/\beta$. Let us 
concentrate on the term corresponding to a given value of $n > 0$. We
perform a further change of variables, setting $y_i = x_i - x_{i-1}$ (with
$x_0=0$), the intervals between the reduced time of flips in the trajectory.
The Jacobian of this change of variables being 1, the integral
over $x_1,\dots,x_{2n}$ can be rewritten as
\begin{widetext}
\begin{equation}\begin{split}
&\int_0^1 \de y_1 \dots \de y_{2 n} \ \I(y_1 + y_2 + \dots + y_{2n} \le 1)
\ \delta(2y_2 + 2y_4 + \dots + 2 y_{2n} - (1 - m \s)) \\
&= \frac{1}{2} \int_0^1 \de S_{\rm o} \de S_{\rm e} \ \rho_n(S_{\rm o} ) 
\rho_n(S_{\rm e} ) \ \I(S_{\rm o} + S_{\rm e} \le 1 ) \ 
\delta\left( S_{\rm e} - \frac{1-m \s}{2} \right) \\
&= \frac{1}{2} \rho_n\left(\frac{1-m\s}{2}\right) \int_0^{\frac{1+m\s}{2}} \de S
\ \rho_n(S) \ .
\end{split}\end{equation}
\end{widetext}
Here we have used $\I(\cdot)$ as the indicator function of an event, 
$S_{\rm o}= y_1 + y_3 + \dots + y_{2n-1} $ and 
$S_{\rm e} = y_2 + \dots + y_{2n}$ the sum of the odd/even $y_i$'s, and 
$\rho_n(S)$ as the density of the distribution of the sum of $n$ independent
random variables uniformly distributed on $[0,1]$. It is easy to prove by
recurrence that for $S\in[0,1]$ one has $\rho_n(S) = S^{n-1}/(n-1)!$. 
Collecting these facts together leads to
\begin{eqnarray}
w_{\rm s}&=& \delta(m-1) + \delta(m+1) 
+ \sum_{n=1}^\infty \frac{1}{n! (n-1)!}
\frac{(\beta B)^{2n} (1-m^2)^{n-1}}{2^{2n -1}} \ ,
\end{eqnarray}
which is indeed equal to (\ref{eq_ws}) thanks to the series expansion of
the Bessel function.


\end{document}